\documentstyle[12pt,aasms4]{article}
\accepted{11 August 2000}
\slugcomment{Astrophysical Journal}

\lefthead{Thompson et al.}
\righthead{Star formation in the HDF}

\begin{document}

\title{STAR FORMATION HISTORY IN THE NICMOS NORTHERN HUBBLE DEEP FIELD}

\author{Rodger I. Thompson}

\affil{Steward Observatory, University of Arizona,
    Tucson, AZ 85721}

\author{Ray J. Weymann}

\affil{Carnegie Observatories, Pasadena CA 91101}

\author{Lisa J. Storrie-Lombardi}

\affil{SIRTF Science Center, California Institute of Technology, 
	MS 100-22, Pasadena CA 91125}

\begin{abstract}

We present the results of an extensive analysis of the star formation
rates determined from the NICMOS deep images of the northern Hubble
Deep Field.  We use SED template fitting photometric techniques to
determine both the redshift and the extinction for each galaxy in our
field.  Measurement of the individual extinctions provides a correction
for star formation hidden by dust obscuration.  We determine star
formation rates for each galaxy based on the 1500 angstrom UV flux and
add the rates in redshift bins of width 1.0 centered on integer
redshift values. We find a rise in the star formation rate from a
redshift of 1 to 2 then a falloff  from a redshift of 2 to 3.
However, within the formal limits of the error bars this could also be
interpreted as a constant star formation rate from a redshift of 1 to
3.  The star formation rate from a redshift of 3 to 5 is roughly
constant followed by a possible drop in the rate at a redshift of 6.
The measured star formation rate at a redshift of 6 is approximately
equal to the present day star formation rate determined in other work.
The high star formation rate measured at a redshift of 2 is due to the
presence of two possible ULIRGs in the field. If real, this represents
a much higher density of ULIRGS than measured locally. We also develop
a new method to correct for faint galaxies or faint parts of galaxies
missed by our sensitivity limit, based on the assumption that the star
formation intensity distribution function is independent of redshift.
We measure the 1.6 $\micron$ surface brightness due to discrete sources
and predict the 850 $\micron$ brightness of all of our galaxies based
on the determined extinction.  We find that the far infrared fluxes
predicted in this manner are consistent with the lack of detections of
850 $\micron$ sources in the deep NICMOS HDF, the measured
 850 $\micron$ sky brightness due to discrete sources and the ratio of 
optical-UV sky brightness to far infrared sky brightness.  From this we
infer that we are observing a population of sources that contributes
significantly to the total star formation rate and these sources are not
overwhelmed by the contribution from sources such as the extremely super
luminous galaxies represented by the SCUBA detections. 
We have estimated the errors in the star formation rate due to a variety of 
sources including photometric errors, the near-degeneracy between reddening 
and intrinsic spectral energy distribution as well as the effects of sampling 
errors and large scale structure. We have tried throughout to give as
realistic and conservative an estimate of the errors in our analysis as possible. 

\end{abstract}


\section{Introduction} \label{sec-intro}

Determination of the history of the star formation rate per comoving volume 
in the Universe is the focus of a great deal of current research.  
Initial studies measuring the UV fluxes of galaxies
indicated a peak in the star formation rate at a redshift near 1.5 
(\cite{mad96}) whereas more recent studies favor a roughly constant star 
formation rate from a redshift of 1.5 back to a redshift of four 
(\cite{stei99}, \cite{saw97}, \cite{pas98}).  
Further support for this conclusion comes from submillimeter
observations of the HDF (\cite{hug98}) and other regions (\cite{sma97},
\cite{barg98}, \cite{ivi2000} and \cite{barg2000}).  The
submillimeter emission is a measure of the UV and optical flux
absorbed by dust in the galaxies and reemitted at longer wavelengths.  In 
fact, uncertainty in the amount of extinction of the UV light is a
major limitation in the use of the 1500 angstrom flux as a measure of  star formation rates in a galaxy.

More recent work by \cite{mad98} indicates that the falloff of the star 
formation rate is not as steep as first thought, but is still present in the 
range of redshifts between 2 and 4.  Madau 1999, however, points out that 
``If stellar sources are responsible for the photoionization
of the intergalactic medium at a redshift of $\approx 5$ then the rate of star formation
at this epoch must be comparable to or greater than the one inferred from 
optical observations of galaxies at $z \approx 3$''. Galaxies with redshifts 
greater than 5 are indeed observed (\cite{dey98}, \cite{wey98}) so that 
reionization due to galaxy starlight may well have occurred before that epoch.
In fact, the recently discovered QSO by \cite{strn00} clearly indicates that 
reionization had already occurred to some degree at a redshift of 5.5.

A key to greater accuracy in measuring the star formation rate history is 
a determination of both the extinction and redshift of the galaxies 
used in the analysis.  \cite{mad99} uses a single correction for dust
absorption of A$_{1500} = 1.2$ mag for all of the galaxies in his sample 
except for the redshift 0.75 to redshift 1.75 galaxies where the equivalent extinction
at 2800 angstroms is used.  \cite{stei99} make corrections for dust absorption 
by assuming that any color deviations in their sample galaxies are solely due to dust absorption, based upon work of \cite{meu99}. These authors derived an 
empirical relation between the slope of the observed
UV spectrum and the far infrared emission to correct their star formation 
rates. In this paper however, we measure the extinction of each galaxy in 
our sample without the assumption of a uniform UV spectral energy distribution 
for all star-forming galaxies.

This work focuses on the portion of the
northern Hubble Deep Field (HDF) covered by deep observations
with the Near Infrared Camera and Multi-Object Spectrometer (\cite{thm99}). 
These observations have the disadvantage of small areal coverage.
However, this portion of the HDF has the significant advantage
of deep photometric images in six wavelength bands, four from the original 
WFPC2 study (\cite{will96}) and two from NICMOS imaging.  The presence of 
the two infrared bands facilitates the technique of photometric redshift 
determination and the extension of the wavelength range to more than a factor 
of 5 provides the opportunity for extinction measurement.

Our analysis of the star formation rate in the deep NICMOS HDF thus consists of two
parts.  The first is a traditional analysis of the star formation rates
via photometric redshift measurements of the galaxies by matching 
numerically redshifted standard spectral energy distribution templates,
but without any correction for internal dust extinction.  This is similar
to the approach used by \cite{fs99} and others.
(A recent summary of various photometric redshift techniques may be found
in \cite{wey99}.)  The second part expands on this by including extinction as well as redshift in the templates in an attempt to reduce the uncertainty in 
the true UV flux of a galaxy caused by dust extinction.

The plan of the rest of the paper is as follows: In section~\ref{sec-obsdat} we 
describe the data set and its photometry. Section~\ref{sec-method} 
contains the methodology we use to measure redshifts, extinction and 
stellar population types. Section~\ref{sec-results} presents the results 
for the redshifts and extinction. In section~\ref{sec-msfrh} we show the 
resultant star formation rates measured from our template fitting. 
These fits are of course subject to both photometric errors 
and template errors.  Section~\ref{sec-err} addresses these issues, including 
the issue of near-degeneracy between extinction and template type.  
Section~\ref{sec-ipo} presents particularly interesting or problematic galaxies.
 In addition to  errors in our fits, the star formation rates at high redshifts 
must be corrected for incompleteness due to both star formation occurring in 
galaxies too faint to be detected and in portions of galaxies whose surface 
brightness is below our detection threshold. This is discussed in section~\ref{sec-csfrh} where we apply a correction both through 
standard luminosity function and aperture corrections, 
as well as by a new method using an empirical form of the ``star 
formation rate surface density distribution function'', recently discussed 
by \cite{lanz99}. Our final numbers for the star formation history versus
redshift are presented in this section.  Section~\ref{ssec-csubfarir} presents 
a prediction of the far infrared fluxes expected for the galaxies in our list.   In section~\ref{ssec-uni} we examine the errors due to large scale structure and small number statistics. We discuss in section~\ref{sec-disc} the comparison 
between our findings and the results from optically based studies, the possible
evolution of the surface brightness of galaxies and the consistency 
between our results and those from submillimeter and far infrared measurements.
Also we comment on the relation of our work to some recent models for star
formation in the early universe. We end with a brief set of conclusions in 
section~\ref{sec-conc}.

We adopt throughout this paper an open Friedman cosmology with
$H_{o} = 65$ and $\Omega_{o} = 0.3$.

\section{Observational Data} \label{sec-obsdat}

The data set for this work comes from the NICMOS and WFPC2  observations of 
the northern HDF.   The NICMOS images used in this analysis are the 
IRAF--generated F110W and F160W images described in \cite{thm99}.

\subsection{Object Selection and Photometry} \label{ssec-pa}

\cite{sza99} describe a method for combining images from several wavebands  
which gives appropriate weights to the various images.
We have followed the Szalay et al. 1999 procedure for selecting galaxies, a 
subset of which comprise the galaxies discussed in this paper. 

The F160W and F110W images were trimmed from their original 
512 x 512 $0.1\arcsec$ drizzled pixel sizes to 481 x 486, eliminating
the outer regions of the frames where the dither pattern caused these regions to
be much noisier than the inner regions. The four  WFPC2
images were then transformed to the same pixel coordinate system as the NICMOS
images by measuring a large number of compact objects and using the IRAF tasks GEOMAP and GEOTRAN. The transformation has an accuracy of about 
0.02\arcsec, with the uncertainties probably dominated by intrinsic differences 
in the centroids for the different bandpasses. Next, we convolved the 
transformed WFPC2 images with a gaussian whose width was chosen such that
the radial profile of the central star (WFPC 4-454, NICMOS 145.0) in the images closely matched the radial profile in the F160W image. While the agreement of 
these resultant radial profiles is not perfect, this procedure should be
adequate for the fluxes used to carry out the template fitting described in this paper.

We next laid down a grid of points where we placed bright artificial  point 
sources  and then used the program SExtractor (\cite{ber96}) to determine the 
background and local $\sigma$ at these grid points.  We then fit these grids 
with 2D Chebychev polynomials to define the background and $\sigma$ at
every point in the 6 frames.\footnote {Because the frames were drizzled and the 
WFPC2 frames were further transformed and convolved, there is some
short scale correlation in the pixel-to-pixel noise: nevertheless, the 
distribution of the pixel values is very accurately gaussian, and it is the 
variance of this distribution that we have measured.}

Each of the 6 frames is then normalized to have zero background and unit 
variance. This must be done {\it locally}, especially for the NICMOS frames, 
since the S/N varies substantially across the image and there are also small 
residual background variations in the reduced images described above.
 From these 6 frames we produced a single (weighted) $\chi^2$ map and selected a
threshold in $\chi^2$ for identifying the pixels in the map that lie above 
that threshold. Since we are interested in extending our analysis to quite high 
redshifts, we gave the F300W and F450W images zero weight, since except for 
very blue low redshift objects these two bands contribute little to the
final selection of objects and at high redshifts simply increase the
noise. In fact, there is very little difference in the maps produced
with weighting equally the F606W, F814W, F110W and F160W images and
assigning zero weight to the F606W image. In the following we use the
maps formed from just the F814W, F110W and F160W images.

We selected a threshold of the Szalay R parameter,  
$R = \sqrt{\chi^2} = 2.3$, as the significance level for ``real'' pixels and 
required 3 contiguous pixels (ie for the SExtractor parameter DETECT MINAREA)
for selecting a preliminary list of objects. With these parameters, SExtractor detected 365 objects. 

We then measured fluxes with SExtractor in an  $0.6\arcsec$ diameter
aperture at the centroid found by SExtractor from the $\chi^2$ 
image. Apertures larger than this admit too much sky for the faint objects and 
increase the risk of contamination from nearby galaxies. For apertures much 
smaller than this, registration and PSF matching become concerns.  The 
determinations of the redshift and extinction for each galaxy described in 
section~\ref{sec-method} utilize these fixed aperture fluxes.  However, in order to estimate the total UV flux associated with each galaxy, aperture corrections 
must be applied. We perform the aperture corrections in two different ways. 
First using the aperture correction method described in \cite{yan98} and 
second with a new method that uses the distribution function of the star 
formation intensity.  

For our template fitting algorithm, we also require an estimate of the $\sigma$'s associated with the 0.6\arcsec\  aperture fluxes. As noted, 
this does not scale simply with the square root of the number of pixels in the 
aperture. We have thus used an entirely empirical estimation by again laying 
down a grid (avoiding the ``significant pixels''), and determining empirically 
the $\sigma$ found for blank sky regions by SExtractor through the 0.6\arcsec\ 
aperture for many locations within each grid point, and again
fitting this grid of $\sigma$'s with a 2D Chebychev polynomial.

Small remaining errors in the local background and/or $\sigma$, may still cause spurious objects to appear.  We therefore imposed the following additional criteria to select a subset of these objects. We required that
all selected galaxies have at least one band with an $0.6\arcsec$ 
aperture signal--to--noise value greater than or equal to 3.5. and at least two 
bands with signal to noise values greater than or equal to 2.5. As 
shown by \cite{hogg98} objects noisier than this are subject to a systematic 
overestimate of their true flux.  There will be a systematic bias in the 
measured flux when the number counts increase with increasing 
magnitude. These authors suggest that objects with signal to noise less than 
$\sim$4 are of little value. We have relaxed this to 3.5 since the actual slope 
of the $log \, N-m$ relation is somewhat shallower than the shallowest
considered by these authors.  In addition, SExtractor provides warning
flags for objects close to the edge of the image.  There were 35 such cases 
and after careful visual inspection we accepted 10 of these as having aperture 
fluxes not compromised by the proximity of the edge.  We also removed the known 
star (NICMOS 145.0) and two faint spurious objects which were associated
with the diffraction pattern from this star.  These additional considerations
reduced the preliminary list of 365 objects to 282 which form the basis
of our analysis.

\section{Methodology} \label{sec-method}

The main output of our analysis is the most likely redshift, extinction 
and intrinsic spectral energy distribution (SED) for each of the galaxies that 
passed the selection criteria described in Section~\ref{ssec-pa}. We do this by 
taking a group of initial template SEDs and numerically altering them over a 
grid of redshift and  extinction.  We then use a minimum $\chi^2$ technique to 
compare the observed fluxes with templates to find the best match.  
Section~\ref{sec-err} discusses the robustness of the technique and the probable errors associated with it.

\subsection{Redshift Determination} \label{ssec-redsh}
 
Our first task in the analysis is finding the redshift for each galaxy.
The paucity of spectroscopic redshifts in the field
 dictates a photometric technique
for the redshift determination.  We chose a template fitting method which
includes interpolation between 6 discrete template spectral energy
distributions.

\subsubsection{Template versus Polynomial Fitting} \label{sssec-tempvs}

Recently there has been relatively good success in the use of polynomial fitting
to determine the photometric redshifts, e.g. \cite{wang98}.  This method fits a
training set of known redshift object fluxes with a polynomial function  of the 
color of the objects. Different polynomials are fit for different color regimes. This technique has an advantage of being independent of any set of assumptions on the actual SEDs of the objects.  For the data considered here we have chosen 
instead a template fitting technique because the number of galaxies with 
spectroscopic redshifts and photometry in these 6 bandpasses is too small at the higher redshifts to be used as a training set.

\subsubsection{Template Properties} \label{sssec-templates}

We draw our templates from three sources.  The first source is the four 
observed SEDs of \cite{cww80} utilized by several authors.  The unreddened
SED of the set of mean SEDs of  \cite{cal94}  provides an additional observed 
template of an active star-forming galaxy (\cite{cal99x}).  
A final and even hotter template is a 50 million year old continuous star 
formation SED calculated from the Bruzual and Charlot models (\cite{bc96}) with 
a Salpeter IMF and solar metallicity. This theoretical 
SED does not have any emission lines, so we have added emission lines from 
H$\alpha$, (O[III]+ H$\beta$) and O[II] by scaling up those associated with the 
Calzetti SED by the ratio of the UV fluxes in the Calzetti and Bruzual--Charlot 
SEDs. Template 6 is substantially bluer than the most recent unreddened SED for 
local star burst galaxies (cf \cite{cal97}). \cite{cal97} shows that stellar
synthesis models evolving older and redder populations must be added to a very 
young population to reproduce the star burst SED. However, we find
instances where our 50 million year old template 
(without any internal reddening) gives a much 
better fit than the Calzetti template, and for this reason we have added 
this last template.  It could well be that at higher redshifts we are seeing 
galaxies that are so young that they have not had a chance to produce an older 
population. An excellent example of this is NICMOS 184.0, (WFPC 4-473) with
a spectroscopically measured redshift of 5.60 (\cite{wey98}).  Template 6 with
no extinction gives a good match to the observed fluxes and reproduces the blue 
F160W - F110W color of the object.  Template 5 with no extinction
provides a poor match and cannot reproduce the blue F160W - F110W color at
the known redshift of the galaxy.  We find many other objects for which an
unreddened template 5.5 or hotter gives the best fit.

\begin{figure}
\plotone{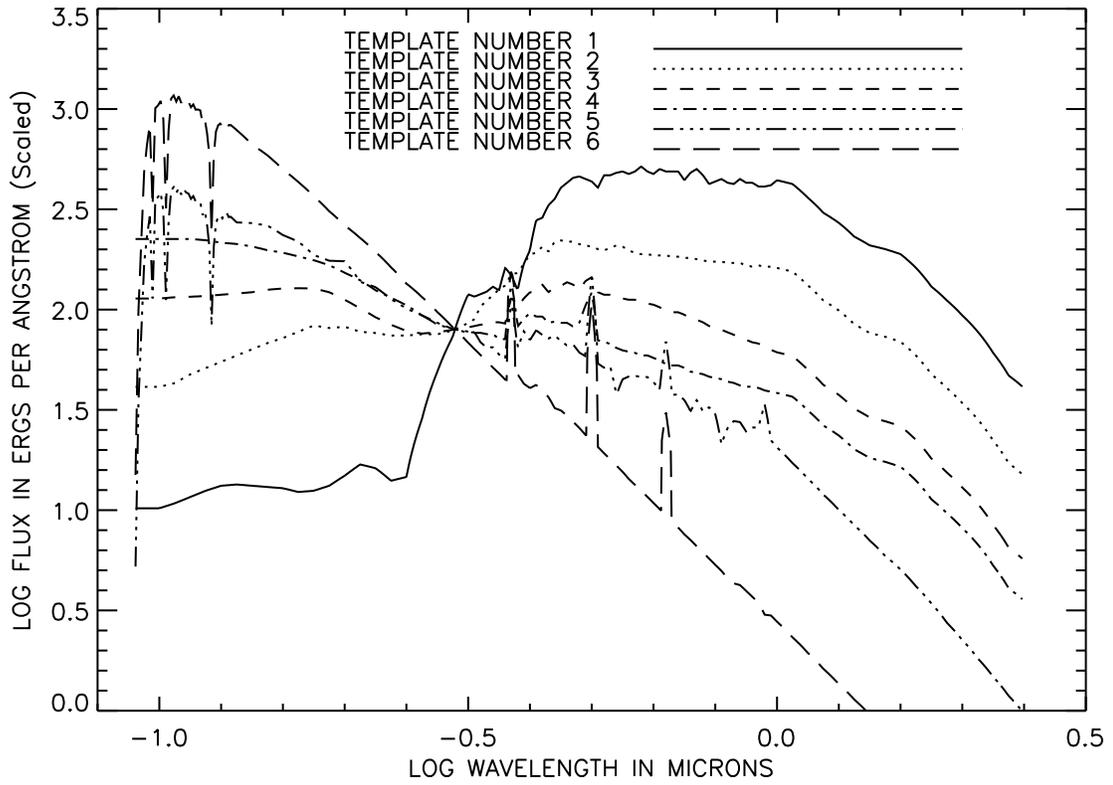}
\placefigure{fig1}
\caption{Spectral energy distributions for the six discrete templates.} 
\label{fig1}
\end{figure}

Figure~\ref{fig1} shows the spectral energy distributions of these six 
basic templates.  The numbers run from the earliest (coolest) galaxy template 
(number 1) to the latest  (hottest) template (number 6). We have assumed that no flux shortward of the Lyman limit escapes from any of the galaxies, and we have 
also neglected any flux in the Lyman $\alpha$ line. Trials with SEDs including 
Lyman $\alpha$ suggest that inclusion of Lyman $\alpha$ makes very little
difference in the obtained fits. In addition to internal reddening from dust 
(described below) we also include the external attenuation from Lyman 
absorption, using the formulation of \cite{mad96}.

\subsubsection{Template Interpolation} \label{sec-templinterp}

The observed galaxies will not be an exact fit to any one of the six templates.
To mitigate this problem and to reduce the extinction error due to an
effect described in Section~\ref{sec-koo}, we interpolate between templates.
The interpolation process is carried out on the total fluxes calculated for
the filters.  Once the filter fluxes for each filter are calculated for all
templates, redshifts and extinctions, 9 intermediate fluxes are calculated
between each template in a linear interpolation.  This effectively increases
the number of templates to 51.  Since the average flux difference between
adjacent templates is about a factor of 3 a linear interpolation should be 
adequate.

\subsection{Extinction Law}\label{ssec-extlaw}

All extinctions used in this work are from the formulation of \cite{cal94}.
Since this extinction law derives from observations of the integrated flux of 
external galaxies it has an advantage of reasonably representing the actual mix 
of scattering and absorption present in real galaxy observations, as well as 
the very complex geometry of the distribution of the hot stars and
dust. The derivation of the ``obscuration law'' of \cite{cal94} for the 
integrated flux from galaxies (and our application of this law) 
assumes an exponential relation between the fraction of the flux
transmitted at any wavelength and the color excess. This implies that the 
geometry of the star and dust distribution more closely resembles a ``clumpy 
screen'' than a  geometry in which the dust and stars are mixed uniformly together.  In the uniformly mixed case the relation between the column density 
of dust and the fraction of stellar radiation escaping takes on a very different form (cf their equation (19)). While it might seem  naive to apply such a law to the integrated colors of an entire galaxy, more recent work (\cite{cal99}) 
using ISO photometry shows that this formulation is able to predict 
reasonably well the average thermal emission (hence the UV extinction) based 
upon the integrated UV slope. As these authors note, however, this result would 
not apply to objects whose geometrical distribution of dust is very different, 
such as very luminous, dusty, compact objects where a large percentage of the 
hot stars are heavily embedded in the dust. For such objects (e.g. the 
``LIRGS'' and ``ULIRGS'') application of the Calzetti formulation is likely to 
lead to a significant underestimate of the fraction of UV radiation absorbed and reradiated by the dust.

We use 15 different extinction values ranging from E(B-V) = 0 to 1.0.
The range between 0.0 and 0.1 is sampled in 0.02 increments and the
range between 0.1 and 1.0 in increments of 0.1.  These extinctions are
applied to each of the 51 filter fluxes of the SED templates to produce
a total of 765 different effective templates at each of the 100 steps
in redshift between 0 and 8.  We choose to limit our redshift range to
values less than or equal to 8 because at redshifts greater than this
only the F160W band will have significant flux.  Flux in only that band
cannot discriminate between a high redshift galaxy and a galaxy with
high extinction.  Also our selection criteria for galaxies will exclude
any galaxy with significant signal to noise in only the F160W band.
This will exclude from our list any objects with redshifts
significantly greater than 8 and very high redshift extremely reddened
objects.

\subsection{Modified $\chi^2$  Analysis}\label{ssec-mcsa}

The basic technique minimizes the $\chi^2$ residuals between the observed 
fluxes and those predicted by the various templates which are numerically
shifted over a grid of redshifts and subjected to internal extinction, measured 
by E(B-V), and the intervening Lyman attenuation.  Our technique varies from 
that of previous workers, (e.g. \cite{fs99})  by altering the
error term in the denominator to include a term proportional to the measured flux as well as the estimated error in the measurement  of the flux.  Our 
$\chi^2$ residual is then given by

\begin{equation} \chi(z,E)^{2} = \sum_{i=1}^{6} \left(\frac{(f_i - A \cdot fmod(z,E)_i)} {\sqrt{\sigma_i^2 + (0.1f_i)^2}}\right)^2 \label{eq:chisq} \end{equation}

In equation~\ref{eq:chisq} the index~i refers to the six fluxes used in this work, $f_i$ is the measured flux and $fmod(z,E)$ is the flux predicted by a 
template at a redshift of z and with an extinction value, E(B-V), equal to E.
Note that this is not a formal $\chi^2$ calculation so the usual quantitative 
probabilities associated with formal $\chi^2$ values are not valid. The 
normalization constant A, is chosen to minimize the value of $\chi(z,E)^{2}$ 
and is given by 

\begin{equation} A =  \sum_{i=1}^{6} \frac{f_i \cdot fmod(z,E)_i} {\sigma_i^2 + (0.1f_i)^2} /\sum_{i=1}^{6} \frac{(fmod(z,E)_i)^2}
	{\sigma_i^2 + (0.1f_i)^2} \label{eq:a} \end{equation}

In equations~\ref{eq:chisq} and~\ref{eq:a}, when the measured fluxes were 
negative, they were replaced by 0.0.  In this form the limit of the expression 
at very low flux levels is the standard form with the formal background $\sigma$ dominating the denominator and at high flux levels it is the flux difference 
between the observations and the model divided by 10$\%$ of the flux instead of 
the $\sigma$.  The rationale for this formulation is that at high flux levels 
the errors in the fit will be proportional to flux since the error will be 
dominated by systematic errors in the flux.  

The actual accuracy of the NICMOS flux levels is higher than this and is 
estimated to be on the order of 1 - 2$\%$.  These higher than usual accuracies 
are due to the nature of the observations.  The deep NICMOS HDF observations are
very highly dithered and are of generally extended objects.  This greatly 
mitigates the interpixel effects described in \cite{lau99}. More generally,
it is our belief that in the case of relatively bright objects, the 
{\it percentage} error in the flux differences should be weighted equally, since all 6 bands contain important information. In any case, our results do not 
appear to strongly depend on the precise coefficient of the flux in the
denominator of equation~\ref{eq:chisq}.

\subsection{Zero Extinction Redshifts} \label{ssec-zer}

The star formation rate results for zero extinction correction plotted
in figure~\ref{fig6} use redshifts that are calculated from a set of
templates that are restricted to zero extinction.  This is a suite of
51 effective templates. In some cases the redshifts can differ from
those calculated with extinction for reasons presented in
Section~\ref{ssec-tgen}.

\section{Results} \label{sec-results}

In this section we present the results of the $\chi^2$ analysis for the 
redshift and the extinction.   Table~\ref{tab1} contains a list of the
results of our analysis on all of the objects, ordered by RA. 
Column 1 of the table gives the NICMOS identification numbers.  Objects with ID
numbers less than 1000 are objects that are identified with sources in
the original KFOCAS catalog of \cite{thm99} and are listed by their original
catalog numbers from the extended electronic version.  Objects 
with catalog numbers greater than or equal to 1000 are SExtractor objects that 
did not  match in position with an original catalog entry to within 
$0.3\arcsec$.  There are several reasons for this.  One reason is different 
morphology in the bands used to determine the position.  The original catalog 
took the F160W band centroid as the position of the object.  The current work 
establishes a position using the weighted sum of the F814W, F110W and F160W 
bands which can shift the location of the centroid.  If the positional 
differences are greater than $0.3\arcsec$ then it is declared a mismatch.  
A second reason involves the difference in the way that the two programs 
determine parent and daughter objects.  There is generally a mismatch in these 
cases.  Finally, our current object selection procedure differs slightly 
from that used in \cite{thm99} so that some faint objects may appear in 
\cite{thm99} which do not appear here and {\it vice versa}.

\placetable{tab1} 

Column 2 gives the WFPC2 identification number of the galaxy from \cite{will96}.
Again there must be a positional coincidence within $0.3\arcsec$ for a match
to be valid. Columns 3 and 4 give the determined values for the redshift
and extinction. Column 5 contains the star formation rate (SFR) in solar 
masses per year as determined in Section~\ref{sec-msfrh}.

The bolometric luminosity of the galaxy is in column 6. The flux of a
galaxy is obtained by integrating over the unextincted selected
template scaled by the factor A from equation~\ref{eq:a}.  The
bolometric luminosity then follows from the redshift and our adopted
cosmology.  The fraction of the luminosity that is extincted and
therefore goes into far infrared flux is given in column 7 followed by
the calculated 850 $\micron$ flux (mJy) in column 8 (cf
section~\ref{ssec-csubfarir}).

Columns 9 and 10 give the template number T of
the best fit, and the modified $\chi^2$ value of the fit from 
Equation~\ref{eq:chisq}. It should be noted that the distribution of the 
modified $\chi^2$ values will not rigorously follow a true $\chi^2$ 
distribution. The values are provided to give a qualitative indication of the
relative goodness of fit for the best fit values of redshift, template type
and E(B-V) from object to object. 

Column 11 gives the total F160W AB magnitude (Tot. mag) derived from
the aperture correction method described in \cite{yan98} followed by
the 0.6 aperture F160W magnitude (Ap. mag) in column 12. If no F160W
magnitude is listed the object had a zero or negative measured F160W
flux. Columns 13 and 14 are the right ascension and declination
positions of the object. The RA listing contains only seconds and the
DEC listing only minutes and seconds.  12$\fh$ 36$\fm$ should be added
to the RA and 62$\fdg$ to the DEC. If a source has a slightly different
RA in this analysis than in the original it may not appear in the same
order as in the original catalog.

 \subsection{Redshifts} \label{ssec-redresults}

\begin{figure}
\plotone{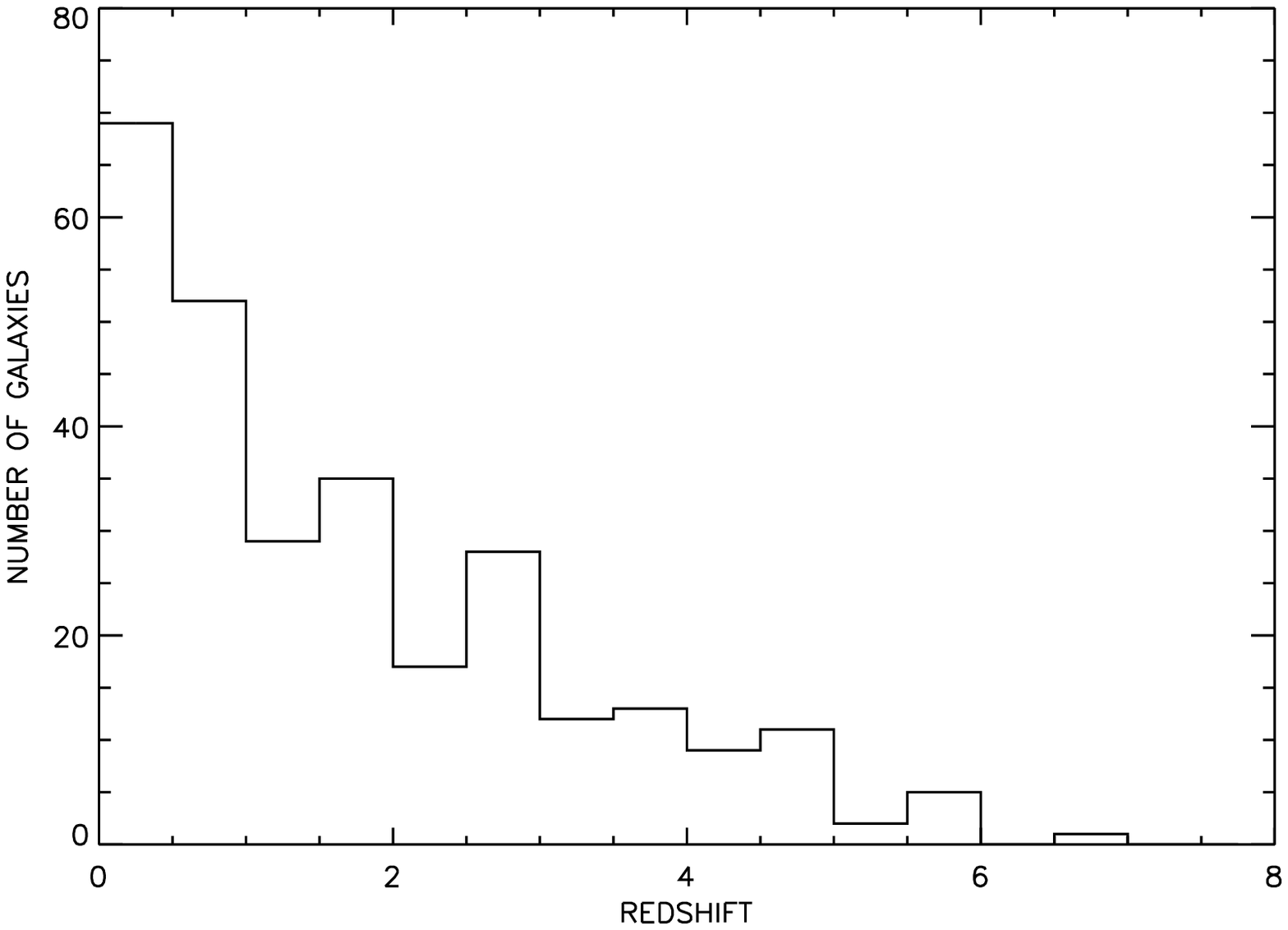 }
\placefigure{fig2}
\caption{Histogram of the number of galaxies versus photometric redshift} 
\label{fig2}
\end{figure}

\begin{figure}
\plotone{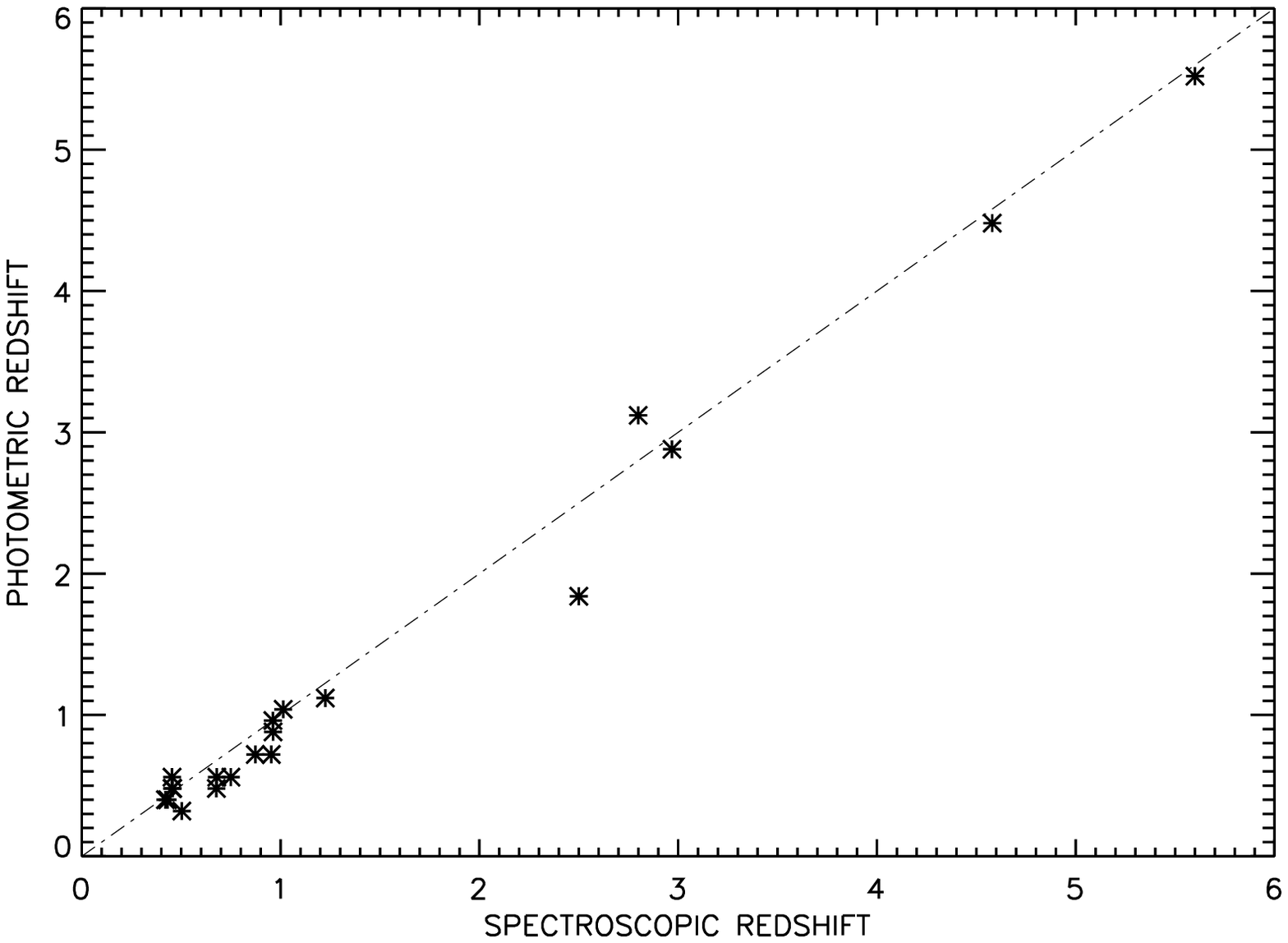}
\placefigure{fig3}
\caption{Comparison of the photometric and known spectroscopic redshifts for 
the deep NICMOS HDF.} 
\label{fig3}
\end{figure}

Figure~\ref{fig2} shows the distribution of photometric redshifts from
our analysis.  A check on the accuracy of our 
methodology and set of templates is a comparison with the known spectroscopic 
redshifts in the deep NICMOS region of the Hubble Deep Field (\cite{coh00}).  
Figure ~\ref{fig3} shows that comparison. Although the number of objects with 
spectroscopic redshifts in the deep NICMOS region is small, the agreement is 
comparable to that typically achieved by the technique of photometric redshifts. (\cite{wey99}).

\begin{figure}
\plotone{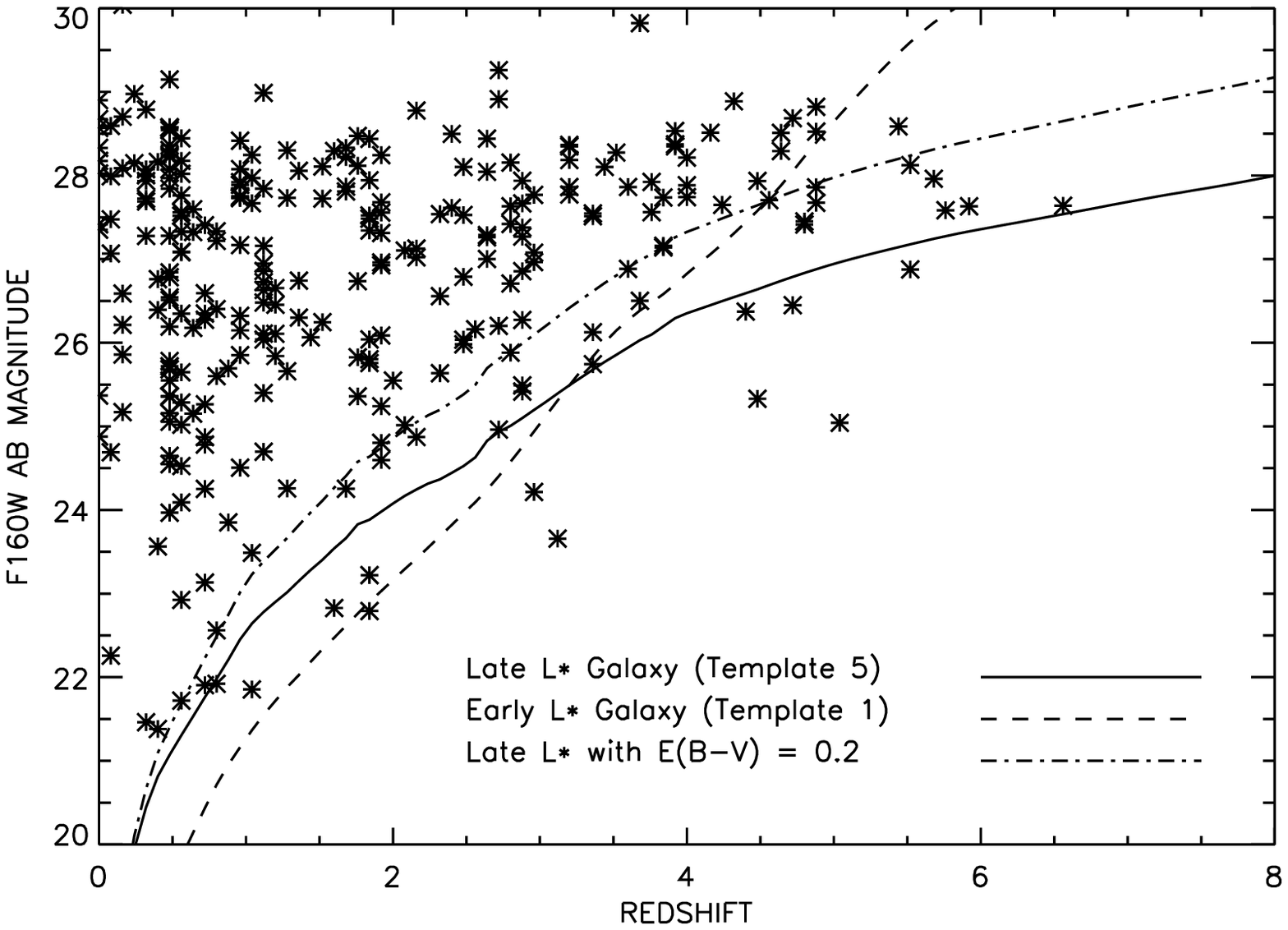}
\placefigure{fig4}
\caption{The distribution of F160W AB 0.6$\arcsec$ diameter aperture 
magnitudes versus photometric redshift. The solid and dashed lines indicate the
F160W aperture magnitude of an early and late-type L$^*$ galaxy of fixed 
luminosity. The dash dot line shows the track of a late-type L$^*$ galaxy with an extinction of E(B-V) equal to 0.2.  The relevance of these plots is 
discussed further in section \ref{ssec-lf}} 
\label{fig4}
\end{figure}

Another check on the reasonableness of our redshift determinations  is the magnitude--redshift plot shown in Figure ~\ref{fig4}.  As expected, the 
brightest magnitude at any redshift dims with increasing redshift.  The plot 
also shows the 0.6$\arcsec$ aperture magnitude tracks of both an early type 
(cool) and a late type (hot) template $L^*$  galaxy in the F160W magnitude 
versus redshift plane.  These are templates 1 and 5 shown in 
Figure~\ref{fig1}. We take the bolometric luminosity of an $L^*$ galaxy to be 
$3.4 \times 10^{10} L_{\odot}$ and assume an exponential profile with a 
characteristic radius of 3.5 kpc. The dashed dot line in Figure ~\ref{fig4} is for an extincted late--type (template 5) galaxy with E(B-V) equal to 0.2

\subsection{Extinction} \label{ssec-extresults}

\begin{figure}
\plotone{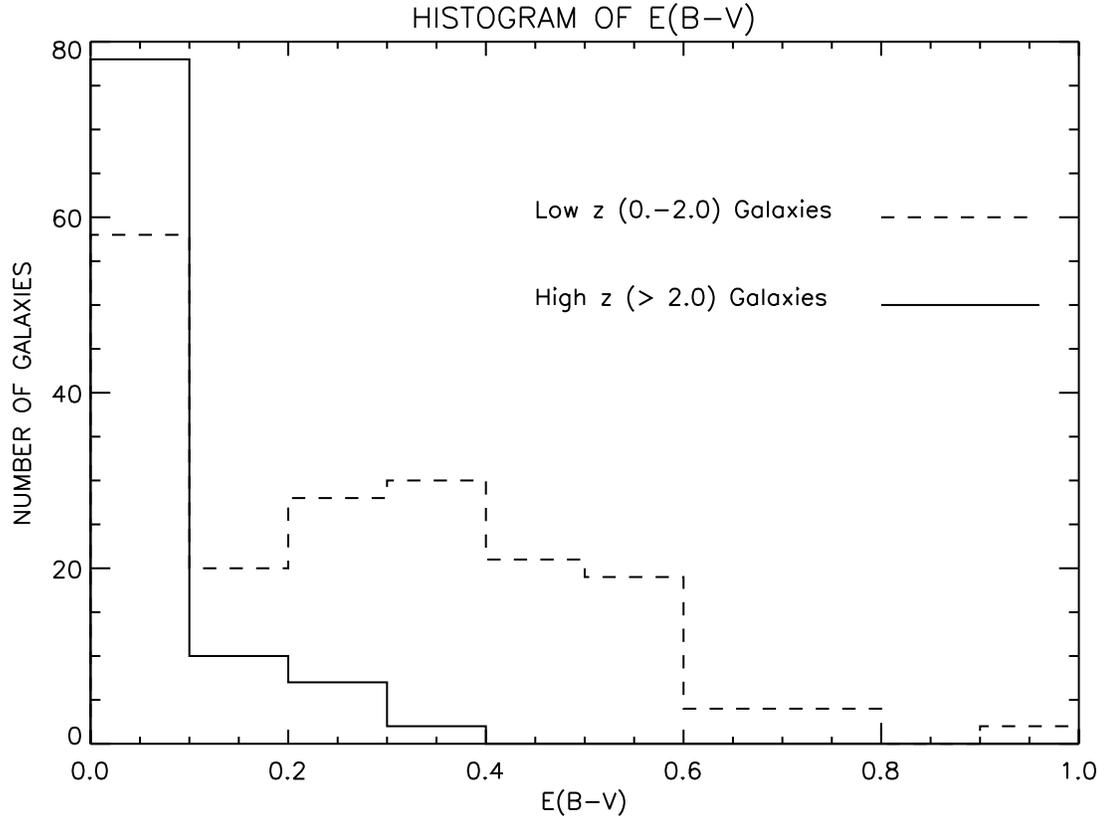}
\placefigure{fig5}
\caption{Histogram of the photometrically determined E(B-V) values.} 
\label{fig5}
\end{figure}

Figure~\ref{fig5} shows the histogram of extinctions. The extinction range is from E(B-V) = 0 to 1. The histograms show significantly more extinction 
for galaxies in the redshift range 0 to 2.0, than for galaxies at higher 
redshift values.  This is not a real effect.  Surface brightness dimming at
the higher redshifts makes the galaxies so faint that high extinction galaxies
fall below our detection limit.  The existence of highly extincted galaxies
at high redshift have been confirmed by SCUBA observations discussed in section~\ref{ssec-scubas}.  The distribution of extinctions in Figure~\ref{fig5}
differs from the distributions of extinction shown in Figure 10 of 
\cite{ade2000}.  In \cite{ade2000} the extinction distribution is based on
the value of the observed UV slope $\beta$ relative to a nominal slope.  
Galaxies bluer than the nominal $\beta$ are assigned negative extinction 
values and the distribution is roughly symmetric about 0.  In our analysis
we determine the best value of the intrinsic UV slope in our template choice
and then only allow positive extinction values to produce the observed UV
slope.  

The 0.0 - 0.1 bin in Figure~\ref{fig5} contains the 5  E(B-V) values of 
0.0, 0.02, 0.04, 0.06, and 0.08, while the 0.1 - 1.0 E(B-V) values are
spaced in intervals of 0.1. There is an overdensity of objects with 
calculated E(B-V) values of 0.0. This may be due to galaxies which 
are bluer than our hottest template.  Even if the galaxy does have some 
reddening the best template match will be our hottest template with no 
reddening.  We have also assigned the \cite{cww80} galaxies zero extinction 
although they must suffer some extinction.  Real galaxies with similar SEDs but 
less extinction will then also be assigned zero extinction in our procedure.

\section{Measured Star Formation Rate History} \label{sec-msfrh}

We determine the observed 1500 {\AA} rest frame flux via the methods described
below.  This flux then determines the observed star formation rate for
the galaxies in this sample. The rates so determined must be corrected
for incompleteness, as described in Section ~\ref{sec-csfrh}.

\subsection{Relation Between Star Formation Rates and the Ultra-Violet Flux}\label{ssec-uvsfr}

This work utilizes the relationship between the star formation rate and the
UV flux at 1500 {\AA} given by \cite{mad98}. 

\begin{equation} UV_{1500} = 8.0 \times 10^{27} \cdot SFR(M_\odot /yr)  \ 
\mathrm{ergs\ second^{-1}\ Hz^{-1}} \label{eq:sfr}
\end{equation}

The 1500 {\AA} UV flux is determined from the redshift, the extinction 
and UV flux of the best fitting template and the scale factor A determined from 
the measured flux of the galaxy (equation \ref{eq:a}). The UV flux of the template is used as a measure of the UV flux of the actual galaxy. We use this 
flux rather than the measured flux since for low redshifts the 1500 {\AA} flux 
is not directly measured and at higher redshifts the F300W and F450W fluxes 
often have relatively high errors.  Also, since the correction for extinction
is a primary goal of this work, we must use the template flux to measure the
intrinsic UV flux in the absence of extinction.  Finally since the Madau UV 
flux to star formation rate is a narrow band relation we must use the template
values to relate the flux in the wide photometric band to the narrow band
UV flux at 1500 angstroms.

The relationship in Equation~\ref{eq:sfr} is dependent on the initial mass 
function (IMF) and therefore may be different at earlier times than at present. 
In particular the IMF may be weighted toward higher mass stars at high redshift when 
the metal content of the star forming material may be less.  This effect would 
produce a higher UV flux for a given star formation rate leading to an over 
estimation of the rate.  This would work in the opposite sense from the
likely underestimate of the star formation rate for the very dusty 
luminous objects mentioned in Section ~\ref{ssec-extlaw}.

\begin{figure}
\plotone{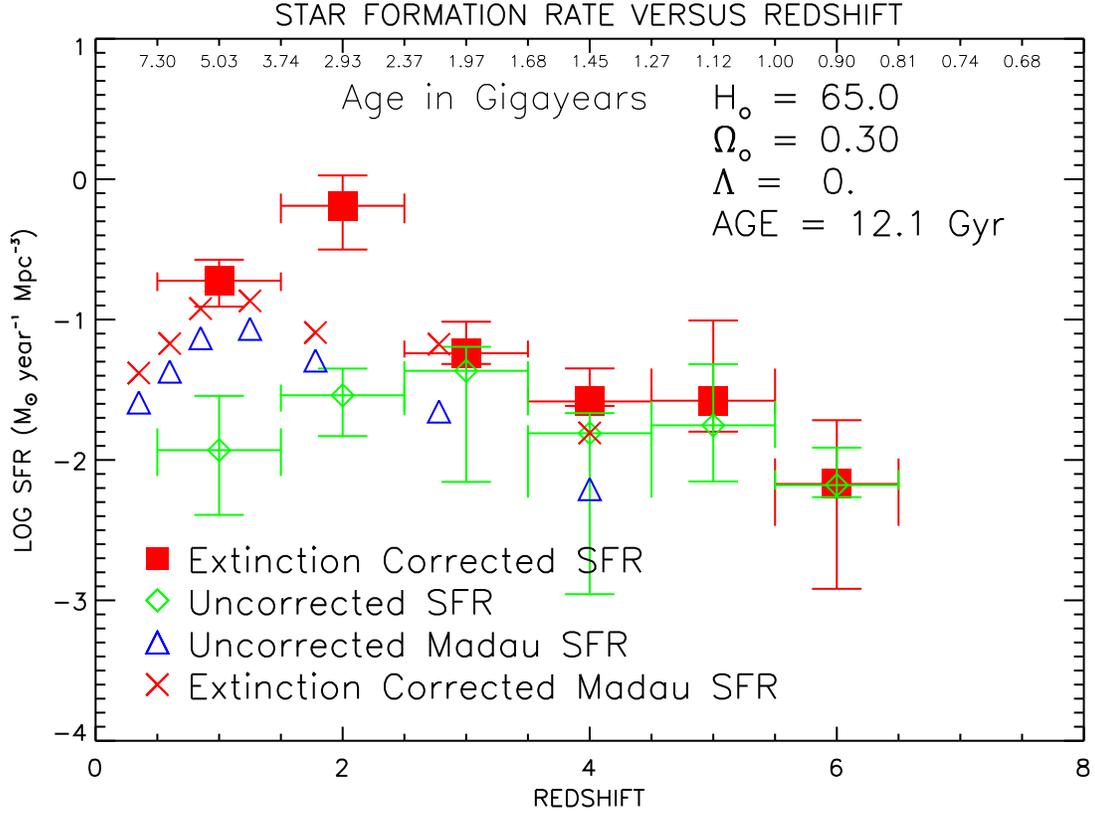}
\placefigure{fig6}
\caption{Star formation rate per comoving volume with and without correction
for the extinction as a function of redshift. The star formation rates have not been corrected for missing luminosity and the error bars reflect only those 
errors caused by errors in photometry. Figure~\ref{fig15} shows the
luminosity corrected star formation rates and full error bars.} \label{fig6}
\end{figure}

Figure~\ref{fig6} shows the combination of all of the analyses from above.
Galaxies with redshifts less than 0.5 are not included in Figure ~\ref{fig6}.
Large area surveys of local galaxies are much more accurate in determining
that result than the small area surveyed here.  The star formation values 
shown in Figure~\ref{fig6} must still be corrected for luminosity missed
due to surface brightness dimming (section~\ref{sec-csfrh}). The error bars in 
figure~\ref{fig6} reflect only those errors caused by errors in 
photometry as discussed in Section ~\ref{sec-err}. Other error sources
are discussed in several following sections and are included in the final
results shown in Figure~\ref{fig16}.

There are several interesting features in Figure~\ref{fig6}.  It is evident
that our measured star formation rate in the 0.5 to 1.5 redshift bin, 
without the extinction correction, is very much lower than the uncorrected rate 
found by \cite{mad99}, as shown by the triangles.  This may be in part due to our selection criteria for the deep NICMOS field.  In order to provide the 
best field for slitless grism spectroscopy we deliberately chose a field that
was the least dense in large, bright and therefore relatively nearby 
objects.  As we will discuss below most of the star formation rate is usually
contributed by a relatively small number of highly luminous objects.  Our 
field selection was biased against precisely such objects.

A second feature is the large correction for extinction present in the
lowest two redshift bins.  This qualitatively follows the trend seen in
Figure~\ref{fig5} where the extinction is significantly higher for
low redshift objects than higher.  Note that our field selection was 
performed on the WFPC2 image so we are not biased against nearby faint, highly 
extincted but intrinsically bright galaxies. Table~\ref{tab2} shows the 10 
highest contributors to the star formation rates in each redshift bin along with their identification numbers from Table~\ref{tab1}. If we look at the 
primary contributors to the increased star formation rate in the lowest 
redshift bin (z = 0.5 - 1.5) we 
find that approximately one third of the rate is provided by one galaxy 
(ID = 49.0) which has an E(B-V) value of 0.5.  29 galaxies out of the 81 in 
the bin contribute fluxes that make up $90\%$ of the total star formation rate,
however, the 3 brightest galaxies contribute more than 50$\%$ of the flux.
For consistency we note that the spectroscopic redshift of galaxy 49.0
is 0.45 which would move it out of our lowest redshift bin while our
photometric redshift of 0.56 moved it into the bin.  We choose not to
adjust by hand those objects with known redshifts.

\placetable{tab2} 

In the second lowest redshift bin (z = 1.5 - 2.5) 5 of 52 galaxies contribute 
$90\%$ of the total corrected star formation rate. One galaxy (NICMOS 166.0)
contributes almost half of the flux and another (NICMOS 277.211) contributes 
another third of the flux.  NICMOS 166.0 and 277.211 have E(B-V) values
of 1.0 and 0.7 respectively and it is the extinction correction applied to
these two galaxies that produces the high star formation rate for this 
redshift bin.  Inspection of the image confirms that both galaxies are extremely
red  with significantly higher near infrared flux than visible.  
Note that the equal extinction correction applied by \cite{mad99} greatly
underestimates the star formation rate for these galaxies. In section~\ref{ssec-csubfarir} we will find that these two galaxies may be
ULIRGs.  The implication of this is discussed in section~\ref{hilum}. 

\section{Error Analysis}\label{sec-err}

Errors in the photometry of the sources and errors due to the 
inadequacy of the standard templates to represent the true spectral energy
distribution of an observed galaxy translate to errors in the values
of the star formation rates in Figure~\ref{fig6}. The first source of error is quantifiable while the second is more difficult to quantify.  

\subsection{Photometric Error} \label{ssec-photerror}

 We test the sensitivity of the calculated redshifts and extinctions
to photometric errors by running 100 test cases on each of the individual 
sources. A single test case consists of randomly altering the flux value in each wavelength band in a Gaussian distribution of errors of width determined by the 
$1 \sigma$ values determined in the photometric reductions plus 1$\%$ of the 
observed flux. This gives 36,500 different cases. Several runs of this 
procedure produced results that were statistically indistinguishable from each 
other so we feel the procedure is robust.  

\begin{figure}
\plotone{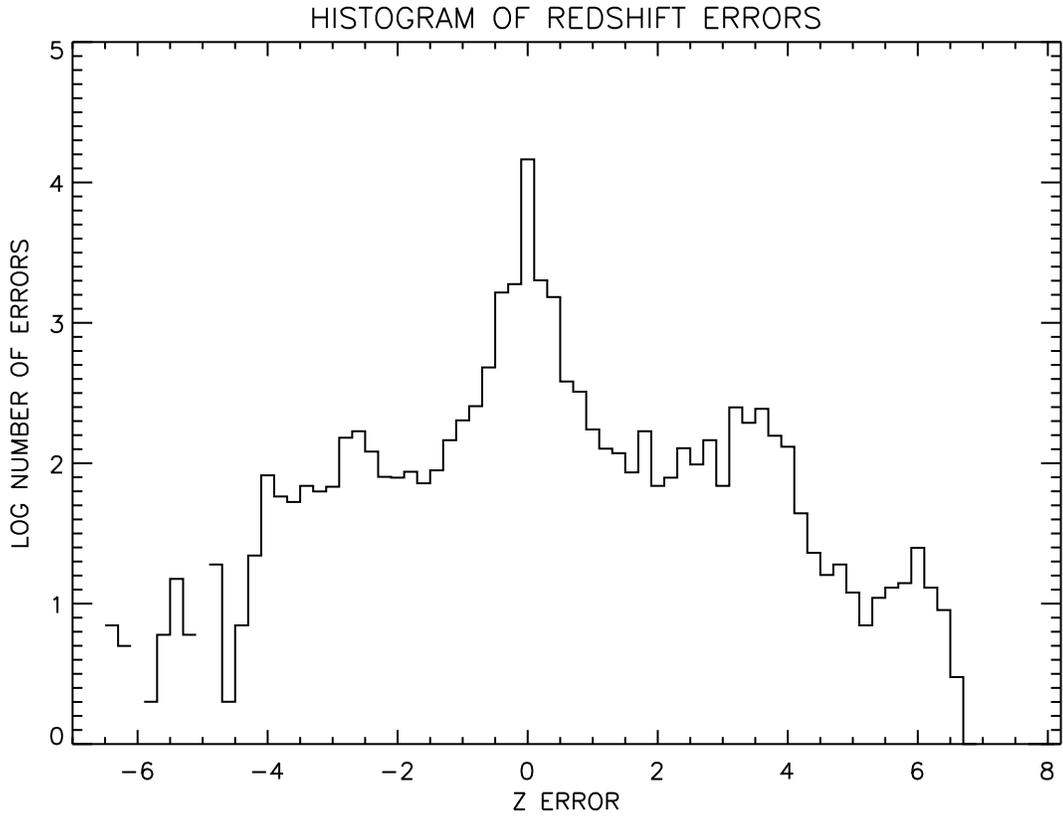}
\placefigure{fig7}
\caption{Histogram of the photometric redshift errors produced from a 
random Gaussian distribution of flux errors.  The bin size is 0.2 in
redshift.} 
\label{fig7}
\end{figure}

\begin{figure}
\plotone{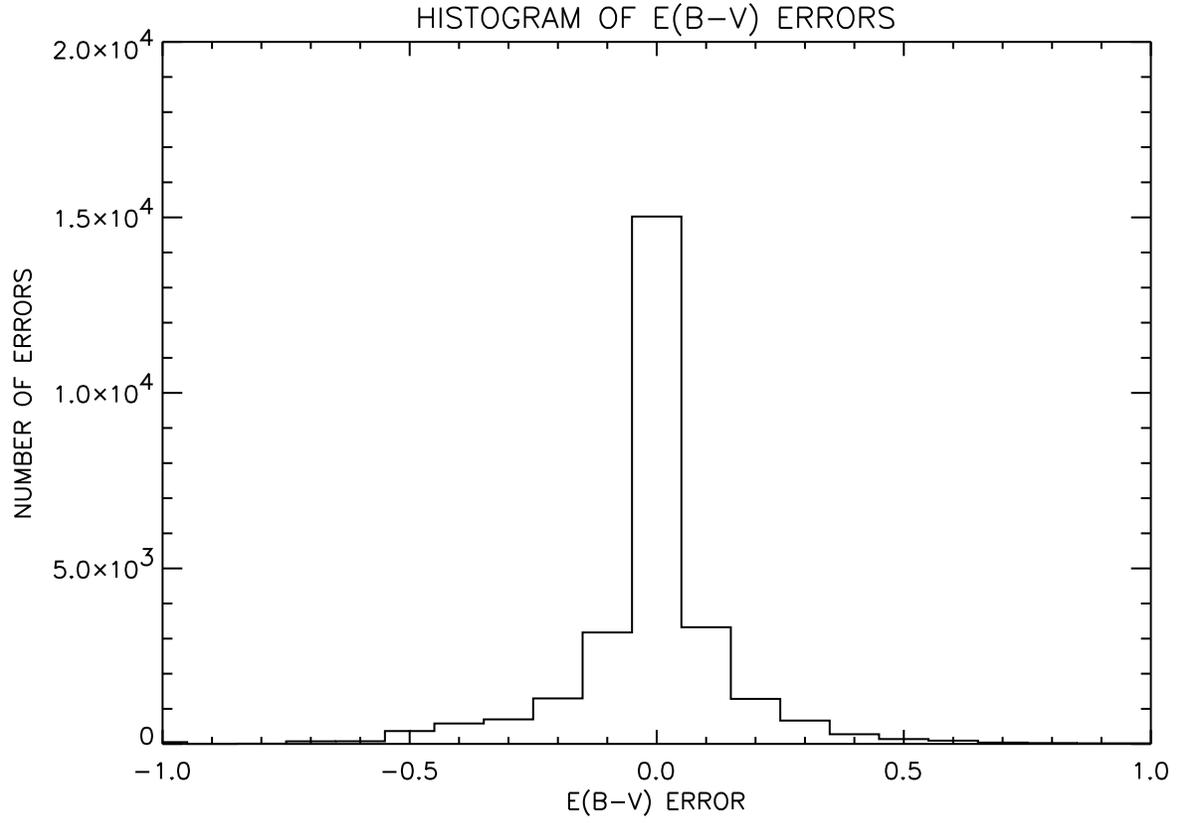}
\placefigure{fig8}
\caption{Histogram of the photometric extinction errors produced from a
random Gaussian distribution of flux errors. The bin size is 0.1 in E(B-V).} 
\label{fig8}
\end{figure}

\begin{figure}
\plotone{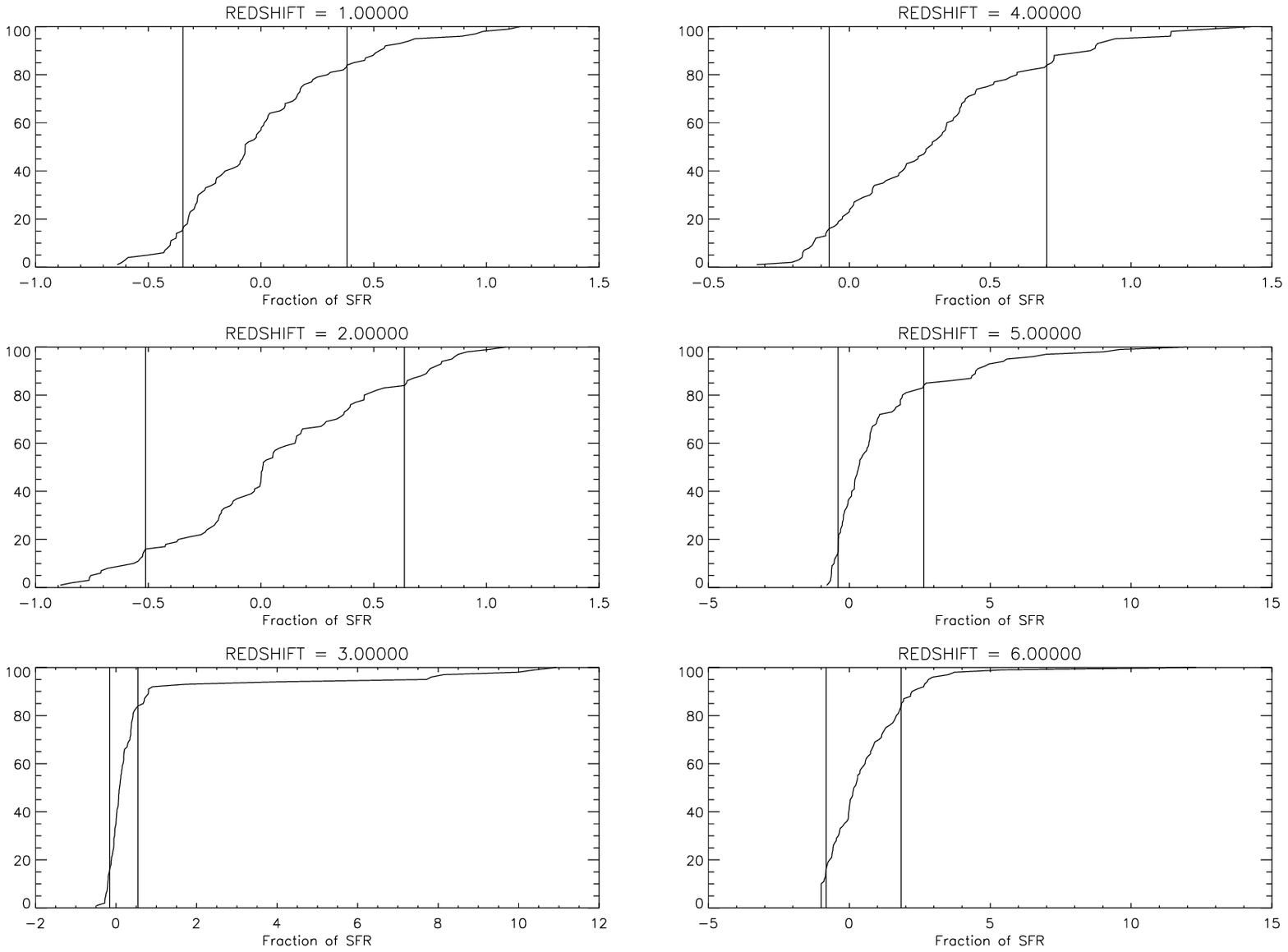}
\placefigure{fig9}
\caption{The cumulative distribution of the fractional errors in derived UV 
luminosity produced from a random Gaussian distribution of flux errors in each
redshift bin.  The left and right vertical lines indicate the 16\% and 84\%
confidence regions. The fractional error is defined as the measured redshift
with perturbation minus the redshift with no perturbation divided by the
no perturbation redshift.} 
\label{fig9}
\end{figure}

Figures~\ref{fig7} and ~\ref{fig8} show the results of 
one run of this procedure for the 36,500 different realizations.
The distribution of errors in redshift and extinction are relatively
symmetric about 0 but the low lying extension to larger errors is not 
consistent with a purely gaussian distribution. Large shifts in
the redshift or extinction occasionally arise when a secondary minimum in the 
$\chi^2$ distribution for the measured fluxes becomes the primary minimum when 
the fluxes are perturbed.

To see how the star formation rates are affected, by the photometric errors, we 
compute the star formation rate in all redshift bins for each
one of the 100 realizations. In Figure~\ref{fig9} we plot these 
distributions along with the 16\% and 84\% values (which would correspond to 
the $\pm 1\sigma$ limits if the distribution were gaussian, which it is not) 
and these values are used for the error bars plotted in Figure \ref{fig6}. 

\subsection{Redshift Errors Resulting from Inadequate Templates} \label{ssec-te}

The error due to improper templates is much harder to quantify than the
errors due to photometry. As shown in Figure ~\ref{fig3}, there
is excellent agreement for the small number of objects in our field
for which spectroscopic redshifts exist. Typical errors
in photometric redshifts for bright objects---bright
compared to the majority of galaxies in our sample---(for which template
errors rather than photometric errors probably dominate) are generally
small compared to the error we have estimated in our set of redshifts
due to errors in the photometry.  See, for example, \cite{ben99}, \cite{bru99} , \cite{bud99}, \cite{con99}, \cite{lanz99}, \cite{wan99}. We therefore neglect
this error compared to the photometric error, though to be sure there
may well be some isolated large excursions where an object at a small
redshift is mistaken for an object at a large redshift or {\it vice
versa} which can give rise to an error in the star formation rate for the
galaxy.

\subsection{Template-Extinction Error} \label{sec-koo}

Too widely spaced discrete templates can give rise to a 
bias  in the determination of the extinction (\cite{koo99}).  Since  negative 
extinctions are not allowed, there is a bias toward a higher calculated
extinction than the actual value. An obvious example is a  galaxy with no 
extinction that lies halfway in color between two templates.  Since negative extinctions cannot be applied to the redder of these two templates, the only way to make a fit is to apply extinction to the bluer of the two templates. Within 
any grid of template and extinction there will be a similar bias toward
higher extinction.  To mitigate this effect, our program uses filter
fluxes interpolated between the filter fluxes from the set of six
primary templates, as described in Section ~\ref{sec-templinterp}.

\subsubsection{Template Extinction Degeneracy} \label{ssec-tgen}

Since extinction and star formation history alter the colors of galaxies in ways which are by no means orthogonal (cf \cite{kod99} and \cite{dthm99})
there can be a partial degeneracy in terms of extinction and template type, leading to a possible error in the derived star formation rate. 
\emph{This error can be present whether the extinction  is explicitly 
determined or not.}  If the extinction is assumed to be zero, a galaxy 
undergoing vigorous star formation but suffering significant extinction may be 
falsely matched to an earlier type galaxy and the actual ultraviolet flux will 
be greater than the flux determined from the match. Conversely, if an extinction correction is attempted, an early type unreddened galaxy might 
be falsely matched with a heavily extincted late type galaxy, particularly in 
the presence of low signal to noise. The large wavelength coverage of our data 
mitigates the problem to a degree, however, this issue must be properly 
investigated. We investigate the issue in several ways, inspection of the 
$\chi^2$ distribution of a heavily extincted galaxy, examination of the 
differences between intrinsically red templates and extincted templates, 
inspection of the effect of photometric errors on the data discussed previously 
in  section~\ref{ssec-photerror}, and finally by performing a similar 
perturbation test on artificial data from our templates.

First,  we chose a galaxy determined to be heavily extincted by our procedure, NICMOS 166, which is also discussed
in Section~\ref{hilum}. The extinction for this galaxy is E(B-V) = 1.0, the highest value in our grid of extinctions. Inspection of the images shows a galaxy that appears relatively faint at optical 
wavelengths but very bright at 1.1 and 1.6 microns.

\begin{figure}
\plotone{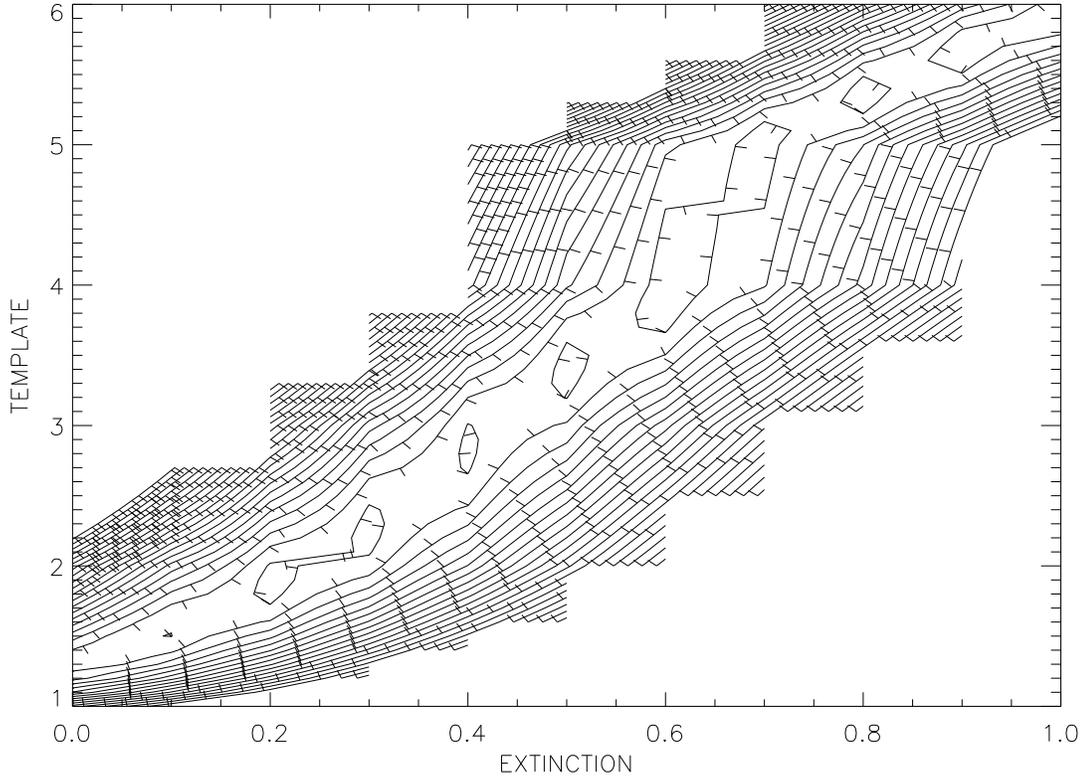}
\placefigure{fig10}
\caption{The contour plot of $\chi^2$ in the template-extinction plane at
the best fit redshift of NICMOS 166.0 (z = 1.60).  Only the lowest contours are 
plotted for clarity.  The tick marks point toward lower values of $\chi^2$.} \label{fig10}
\end{figure}

\begin{figure}
\plotone{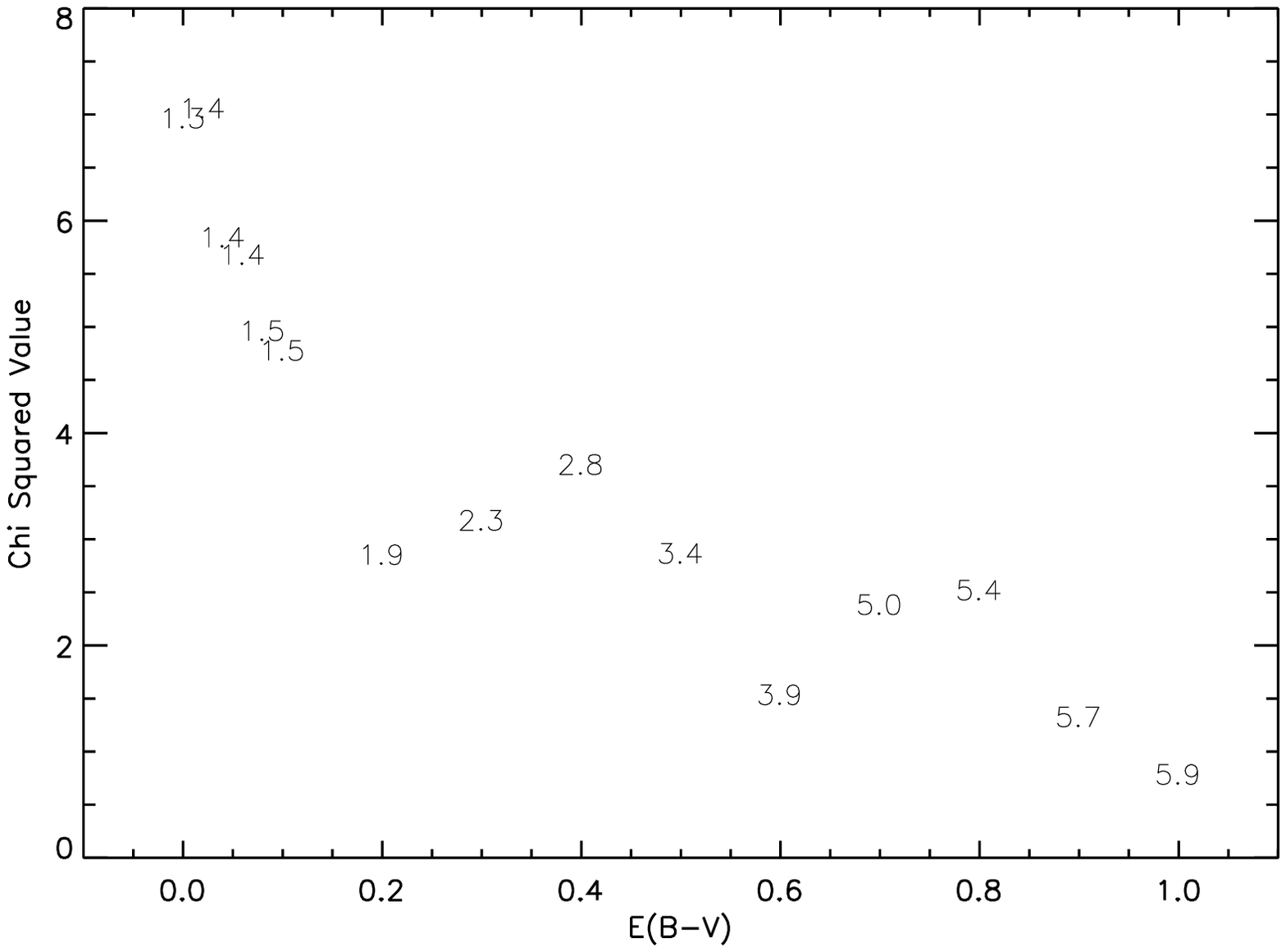}
\placefigure{fig11}
\caption{A plot of the minimum modified $\chi^2$ value at each
extinction for NICMOS 166.  The numbers in the plot indicate the
interpolated template value at each of the minima.} 
\label{fig11}
\end{figure}

The effect is shown in Figure ~\ref{fig10} which displays the contour
plot of the $\chi^2$ map  for the galaxy NICMOS 166 in the template -
extinction plane at the best-fit redshift of the object (z = 1.60).
There is a clear valley of low $\chi^2$ values running from an
early-type SED and low extinction at the lower left to a late-type SED
and high extinction at the upper right. To look in detail along the
line of minima we plot  in Figure~\ref{fig11} the minimum $\chi^2$ for
each of the 15 different extinctions along the track in Figure
~\ref{fig10}.  The numbers refer to the interpolated template number
with 1.0 being the earliest template and 6.0 the latest.  This plot
reveals that there is a significant difference in the modified $\chi^2$
value along the minimum $\chi^2$ track.  The selected very blue late
type template with high extinction is significantly more likely than an
early type template (e.g. elliptical) with low extinction.

\begin{figure}
\plotone{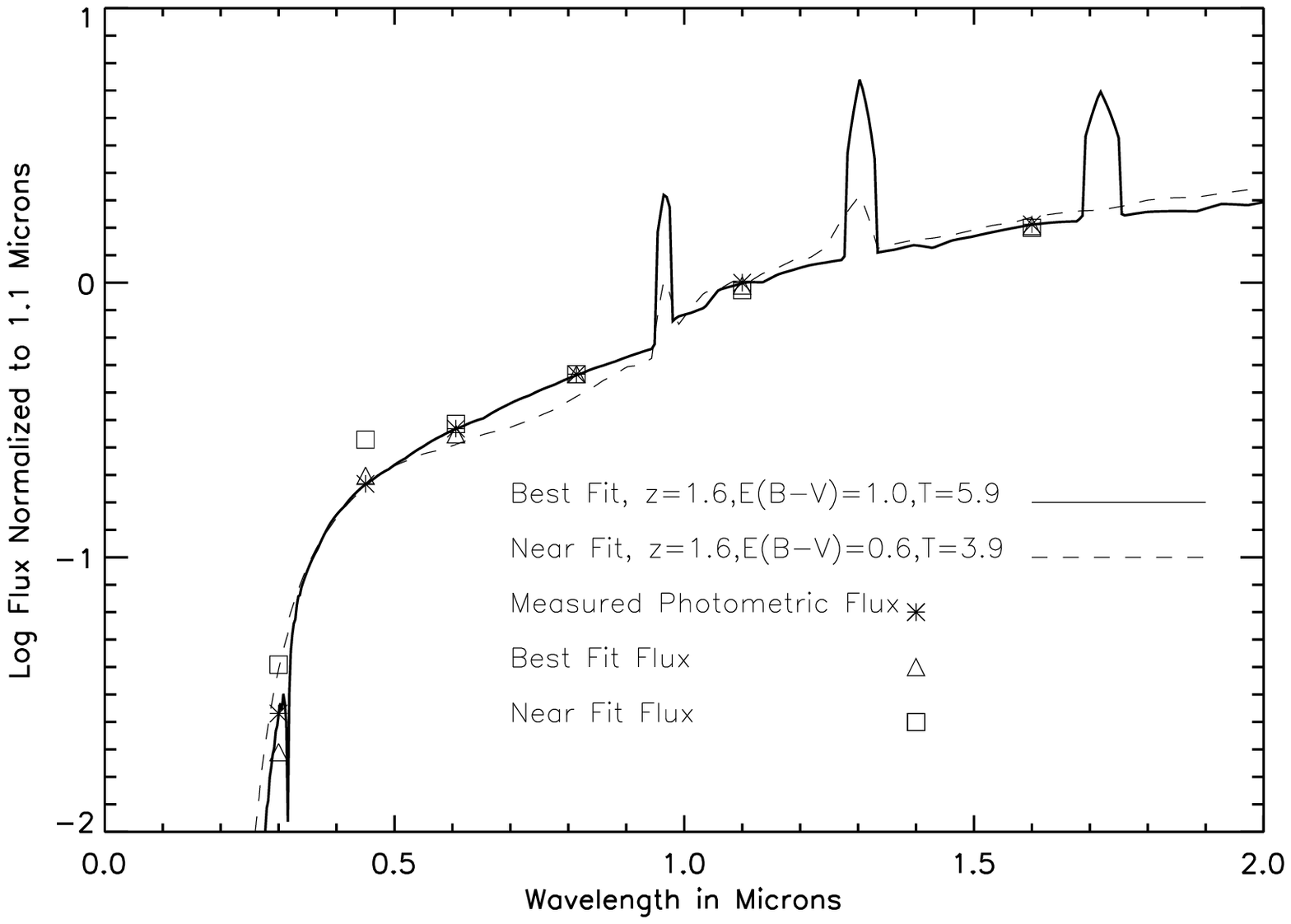}
\placefigure{fig12}
\caption{A comparison between an early (3.9) and a late (5.9) template fit to
the galaxy 166.0} 
\label{fig12}
\end{figure}

To see the reason for the difference in the modified $\chi^2$ value and
to illustrate the near degeneracy we plot in Figure ~\ref{fig12} the
selected SED for the galaxy (template 5.9, E(B-V) 1.0), and a nearly
degenerate much earlier type SED (template 3.9, E(B-V) 0.6).   All have
been normalized to the F110W flux. The measured fluxes of the galaxy
are given by the asterisks, the best template fit by the triangles
 and the near degenerate fit as squares.  Although the best fit has a
$\chi^2$ value of 0.69 and the nearly degenerate fit has a larger
$\chi^2$ value of 1.44, the differences in the fit are very small and
involve mainly the U and B pass bands.

As a second approach towards investigating the reddening--template degeneracy we
imagine  template 5.0 with zero extinction, shifted over redshift space, to
represent the ``observed'' fluxes of galaxies, and then find the best-fitting 
values of redshift and extinction for template 6.0 to match these fluxes. We 
scaled the fluxes of template 5 so that the F160W s/n was 10.0. We chose 
templates 5 and 6 for this exercise since the vast majority of our fits involve 
template types in this range. We find that the best fitting redshift for 
template 6 tracks the input redshift of template 5 very well, with the 
dispersion in $\Delta z/(1+z)$ being less than 5\% over the redshift range 
between 1.0 and 5.5; we get the redshift right even with the template and reddening uncertainties.
The typical value of E(B-V) in the template 6 fit is about 0.17, corresponding 
to an attenuation of about 1.5 magnitudes at 1500 {\AA}. This of course has a 
strong effect on the derived star formation rate. However, if we examine the 
color differences between template 5 and the best-fitting template 6 we find 
that (except for the cases where the fluxes are essentially zero so that the s/n is very low)  these differences are typically only between about 0.1 and 0.2 
magnitudes. Thus, in order to have an accurate determination of the extinction 
and hence the star formation rate we need photometric errors smaller than this 
and  confidence that our set of templates and  reddening law represent real 
galaxies to this same degree of accuracy. The differences will be even less for 
an unreddened template of type intermediate between 5.0 and 6.0 and in fact, 
some objects undergoing vigorous star formation might even be bluer than our 
hottest template 6.

In section~\ref{ssec-photerror} we presented estimates for the effect of 
photometric errors on the derived star formation rates based upon the cumulative distribution function of 100 Monte Carlo simulations of these errors. To first
order this tests the robustness of the method since the perturbations slide
the solution along the template/extinction minima.  A more robust test,
however, as suggested by an anonymous reviewer is to start with artificial 
data that spans all of our template types but with relatively low extinction 
values and then see if the introduction of perturbations systematically drives 
the solution to different star formation rates.

Our artificial data set consists of all of our 51  interpolated filter
fluxes from template types with extinctions restricted between E(B-V)
values of 0 to 0.1.  This provides maximal opportunity for
perturbations to push the solution toward higher extinction, bluer
intrinsic spectrum and hence higher star formation rate.  This data set
contains 306 artificial galaxies from the 51 effective templates and the 6
extinction values spaced in increments of 0.02 from 0. to 0.1.  We pick
the redshift and brightnesses from the redshift 2 bin sources in our
sample along with the 1 $\sigma$ error values described in
section~\ref{ssec-mcsa}.  This should give a realistic distribution of
signal to noise values in a redshift bin where the correction for
extinction is high.  Since there are many fewer sources in the redshift
2 bin than artificial sources, the brightness, redshift, and $\sigma$
values are used between 3 and 4 times but each use is for a different
template type and extinction.  As with the source flux perturbations
each artificial galaxy receives 100 different perturbations to each of
its fluxes.  The first perturbation is zero to establish the true star
formation rate for the ensemble of sources. As expected the zero
perturbation returns the correct values for all of the input
parameters.

\begin{figure}
\plotone{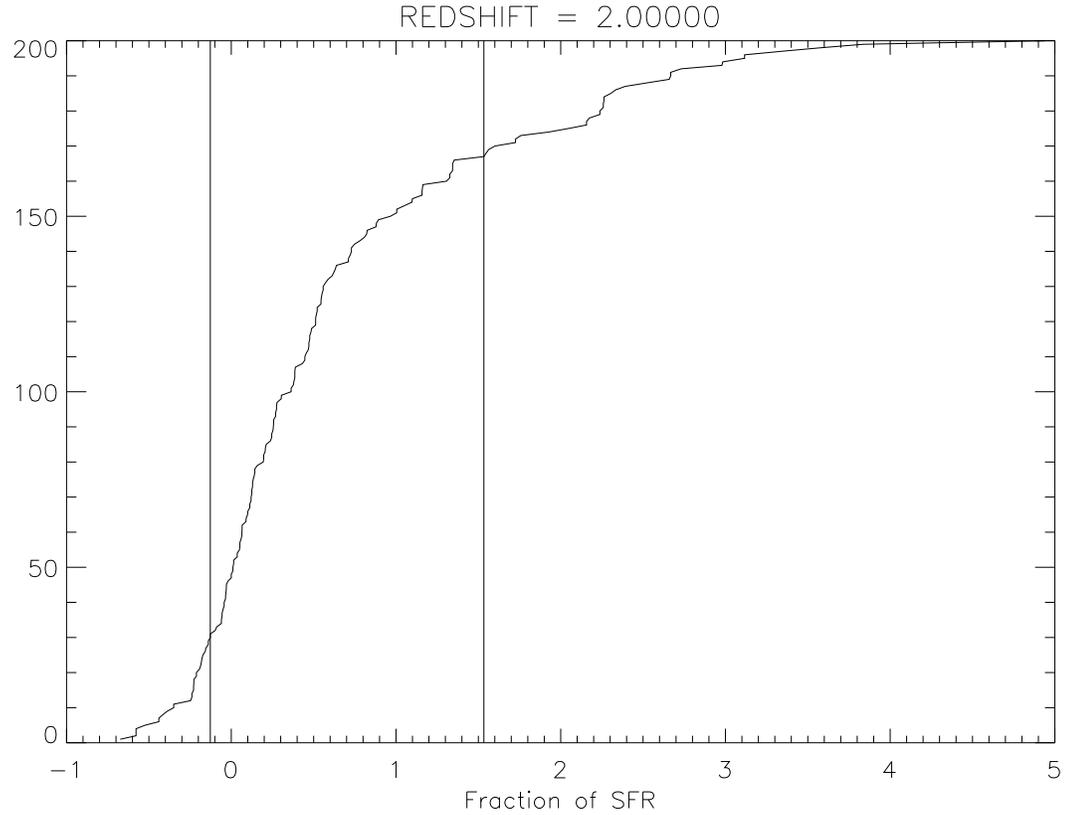}
\placefigure{fig13}
\caption{The cumulative distribution of the fractional errors in derived UV 
luminosity produced from a random Gaussian distribution of flux errors imposed
on artificial data.  The brightness and 1 $\sigma$ errors are from the source
redshift 2 bin.  The left and right vertical lines indicate the 16\% and 84\%
confidence regions.} 
\label{fig13}
\end{figure}

The results are shown in figure ~\ref{fig13} in the same manner as for the
source photometric flux perturbations in figure~\ref{fig9}.  It is evident
from the figure that the range of errors in the output is within our previously
estimated error bars but that there is a systematic trend toward increased
star formation rate.  The net increase is approximately $70\%$ in this
artificial sample.

Although the results of this analysis provide a factor that might be used to
revise our values of star formation rates derived from our method, the true
factor is probably less than the one derived here.  The input sample was of
course set up with no high extinction galaxies so there was little error space 
on the negative side of the perturbation analysis.  The actual sample 
has several galaxies with derived extinctions significantly greater
than 0.1.  Also the brightest galaxies dominate the star formation rate
and it is for these galaxies that we have the highest photometric accuracy.
We could divide all extinction corrections by 1.7 or less, but, since the
actual number is uncertain we will assign a factor of 2 error due to this
effect with the knowledge that this probably over estimates the lower error
bar. Changes of a factor of 2, however, do not alter the conclusions of this work.

Ultimately, it would be preferable to utilize
FIR or submillimeter measures in addition, but this prospect is still quite 
distant at the flux levels we are dealing with, as elaborated upon in section 
9.5. We hasten to add that in previous work where no account was taken of 
internal dust extinction, the true star formation rate will of course be 
systematically underestimated.
\section{Interesting and Problematic Objects}\label{sec-ipo}

There are several galaxies in our field that display interesting 
characteristics. In this section we draw attention to some of these, partially 
in the hope that follow up observations may shed additional light on the nature 
of these objects.

\subsection{Possible Very High Redshift Object}

There is a single object for which we derive a photometric redshift greater than 6, NICMOS 118.0, at Z = 6.56.  This is a relatively faint object, but with a 
significant amount of flux in both the F110W and F160W bands (with formal S/N above 6 in both bands) with a relatively blue F110W - F160W color. It has a 
s/n  through our 0.6\arcsec\ aperture of less than one in the F814W and F606W 
bands. This is, therefore, on the face of it, an excellent candidate for a very 
high redshift object. However, detailed inspection of the original WFPC2 F814W 
and WFPC2 606W images clearly indicates some flux is present in about 
equal amounts in these two bands, contrary to what is expected for a redshift 
this high. It is thus possible that it is a highly reddened object at a much
lower redshift or perhaps a superposition of a very faint foreground low 
redshift object on a truly high redshift galaxy. Further discussion of this 
object is contained in a forthcoming paper dealing with compact and high 
redshift objects in this field (\cite{lisa00}).  It is, in any case, a very 
interesting object, but unfortunately probably too faint for spectroscopic 
follow up with existing facilities.  Given the problematic nature of this object we do not include it in our analysis of the star formation rate history.

\subsection{High Redshift Early Type Galaxies} \label{early}

Our analysis also shows two objects, NICMOS 165.0 and NICMOS 277.212, with high redshifts (4.40 and 5.04) and relatively early type templates (3.4 and 3.5).  Note that NICMOS 277.212 is clearly a separate galaxy from other indicated ``daughters'' of NICMOS 277.10, a large spiral galaxy. The original KFOCAS 
analysis found that their isophotal areas touched so NICMOS 277.212 was 
indicated as a daughter.  The SExtractor analysis listed it as a separate 
object. Careful examination of the images indicates a large region of diffuse 
flux around NICMOS 165.0 in the F160W and F110W bands which, 
if part of the galaxy, would argue 
against such a high redshift due to the large size.  The contribution of this 
diffuse flux to the 0.6 arc second aperture flux, however, is negligible and 
does not influence the photometric redshift determination.  Similarly with 
NICMOS 277.212 there is diffuse flux from the neighboring galaxies which is why 
it was originally considered a daughter object in \cite{thm99}.  Again the 
contribution of this diffuse flux in the aperture magnitude is not large enough 
to alter the photometric redshift determination.  It is interesting, however,
that at 1.6 microns there is some overlap between 277.212 and the suspected
ULIRG 277.211 which is discussed Section~\ref{hilum}.
If both our photometric redshifts and template types are correct for these two objects, our adopted cosmological parameters indicate that these 
galaxies are approximately 1 gigayear old. The template type, however, implies 
that star formation has been occurring for nearly this same length of time. Thus, these two objects would represent relatively high redshift objects having
intermediate age stellar populations.

\subsection{Starburst and High Luminosity Galaxies \label{hilum}}

Five objects (NICMOS 49.0, 102.0, 166.0, 277.212, 277.211) have
luminosities that classify them as possible Luminous Infrared Galaxies,
LIRGs or Ultra Luminous Infrared Galaxies, ULIRGS.  At the same time
they also have the highest star formation rates among the galaxies.

Two objects, NICMOS 166.0 and NICMOS 277.211, have the highest star formation rates (534 and 375 M$_\odot$ per year) of all the objects in our list.  
They are also highly extincted. The validity of this extinction for 
NICMOS 166.0 is discussed in detail in section~\ref{ssec-tgen}.  If the high
extinction, template choice and redshift of these galaxies are correct
then both of these galaxies have luminosities of slightly
more that $1.0 \times 10^{12} L_\odot$ which makes them ULIRGS. 

\begin{figure}
\plotone{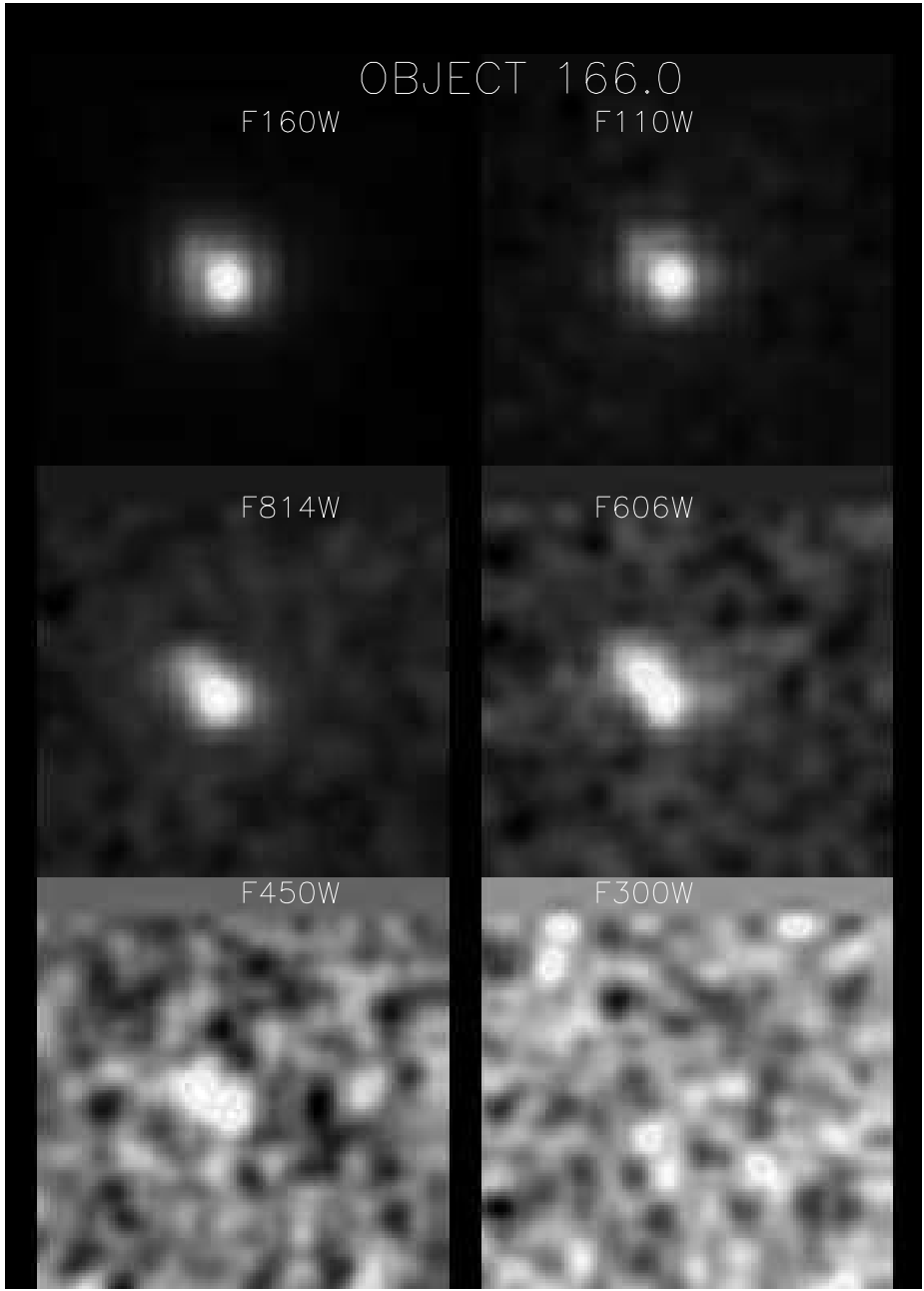}
\placefigure{fig14}
\caption{The image of NICMOS 166.0 in each of the six bands.  All images have been stretched linearly so that the maximum flux is 1 and the minimum flux is 0. This hides the very large decline in flux with decreasing wavelength.} 
\label{fig14}
\end{figure}

At 1.6 microns 
NICMOS 166.0 appears to have a single bright nucleus that is heavily
obscured since it decreases rapidly in brightness with decreasing wavelength.  At optical wavelengths where the nucleus does not dominate the image there 
appears to be an asymmetrical extension while at 0.3 microns the galaxy 
essentially disappears as shown in figure~\ref{fig14}.  

NICMOS 277.211 is in the region of the NICMOS 277.10 complex but is again
clearly a separate galaxy.  Its appearance is similar to NICMOS 166.0 with
a single bright nucleus at 1.6 microns which decreases in intensity with 
decreasing wavelength.  Low surface brightness structure is evident at 
the shorter wavelengths but unlike NICMOS 166.0 it does not disappear 
completely at 0.3 microns.

NICMOS 166.0 has a reasonably secure redshift determination with the
$\chi^2$ value rising sharply for redshifts less than 1.66 so it is unlikely
to be a low redshift and hence a low luminosity galaxy. There is a 
secondary minimum at higher redshift which would of course increase its luminosity.  If the secondary minimum in the template extinction
discussed in Section~\ref{ssec-tgen} is used the luminosity is
decreased to $1.4 \times 10^{11}$ L$_\odot$ which reduces it to LIRG
status.  NICMOS 277.211 has a secondary minimum in its $\chi^2$
distribution at a lower redshift of about 1.4.  Reduction of its
distance from a redshift 1.84 to 1.4 yields a luminosity of $4.4 \times
10^{11}$ L$_\odot$ which would remove it from the ULIRG classification
and demote it to the LIRG classification.

We also note that NICMOS 166.0 and 277.211 are coincident with the ISO
sources PS3 17a and PS3 15 from \cite{aus99}.  The observed flux levels
at 15~$\micron$ are consistent with the fluxes predicted from our dust
model.  This is further evidence for these sources having significant
extinctions.

\cite{ade2000} give an empirical relation between the 850 $\micron$ flux and
the 20 cm flux which is a function of redshift.   With this relation and the
predicted $850 \micron$ fluxes from Table~\ref{tab1}, the predicted 20
cm fluxes for NICMOS 166.0 and 277.211 are 30 and 19 micro Janskys. 
Comparison of the dust spectral energy distribution of \cite{ade2000} and our 
distribution shows that our predicted $850 \micron$ fluxes are roughly a 
factor of two below the \cite{ade2000} fluxes.  In fact the
empirical function has very significant width which would allow the fluxes
to vary by factors of two to three and still be within 1 $\sigma$ of the 
prediction. \cite{mux2000} do not find any detections at the locations of the 
galaxies to a limit of 27 micro Janskys at 1.4 gigahertz.  Radio maps
very kindly provided by Dr. Muxlow in advance of publication show no
contours for NICMOS 166.0 but some contours at the location of NICMOS
277.211 but still below their accepted lower limit.  

If both of these objects are ULIRGs then their presence in one 50 arc second by 50 arc second field is surprising.  The local space density of ULIRGs with
luminosities of $10^{12} L_\odot$ is $8 \times 10^{-8}$ Mpc$^{-3}$ Mag$^{-1}$ 
(\cite{soif87}) which is roughly equal to the space density of quasars.  At 
this density we would expect $1.5 \times 10^{-4}$ ULIRGs in our redshift 2 bin.
The presence of two ULIRGs in our field may indicate a much higher rate of 
merging and hence high luminosity galaxies at a redshift of 2 as suggested in 
some hierarchical galaxy formation models (eg \cite{bln99}).  Given all
of these considerations we chose to designate these galaxies as possible
ULIRGs.

The objects NICMOS 49.0, 102.0, and 277.212 have star formation rates
of 69, 59, and 12 $M_\odot$ per year.  The early classification of the
SED for NICMOS 277.212, discussed in Section~\ref{early}, puts it in a
lower star formation rate category.  All three of the galaxies are at
the border of the LIRG luminosity classification with luminosities just
over $10^{11} L_\odot$.  Both NICMOS 49.0 and 102.0 have most of their
luminosity reradiated as far infrared flux and can therefore be
classified as LIRGS.  NICMOS 277.212 has only 22$\%$ of its luminosity
in the far IR and should be classified simply as a luminous galaxy.
Our dust SED model predicts that NICMOS 49.0 should be an ISO source
and it does correspond to the ISO source PM3 15 of \cite{aus99}.

\section{Incompleteness Corrections to the Star Formation Rate History} \label{sec-csfrh}

We approach the correction to the measured star formation rate history
due to galaxies which are too faint to be detected as well as to
those portions of detected galaxies having surface brightness
below our detection limit in two ways.  The first, and traditional way 
integrates an assumed luminosity function to determine the contribution from 
unobserved objects, using the aperture corrections described in 
Section~\ref{ssec-pa} to account for the outer lower surface brightness portions
of galaxies. The second method uses the observed distribution of star formation 
intensity at low redshifts to correct for the unobserved portion of the 
distribution at high redshift. 

\subsection{Corrections Based on the Luminosity Function and Aperture Corrections} \label{ssec-lf}

It is readily apparent by inspection of Figure~\ref{fig4} that at high
redshifts we are sampling only part of the galaxy luminosity function.  
We can get an estimate of the possible error in our star formation rate by 
asking what portion of the total luminosity we are missing.  The solid and 
dashed lines in Figure~\ref{fig4} represent the expected 0.6\arcsec\ diameter 
aperture magnitudes for an L$^*$ late type galaxy and an L$^*$ early type  galaxy respectively, (templates 5 and 1) with the assumptions described in 
section~\ref{ssec-redresults}. The line for the late type L$^*$ galaxy shows 
that it is easily detectable even at a redshift of 8. If the luminosity 
distribution of galaxies is represented by a Schechter function with a slope of 
$\alpha = -1.6$ then we can integrate the luminosity function for magnitudes 
fainter than our observed F160W AB magnitude limit of approximately 28.5 to 
estimate the missing luminosity assuming a given template type. Since
most of the higher redshift galaxies are best represented by templates 5 or 6
we will use the template 5 line shown in the figure for the calculation. 
The fraction f of missing luminosity is 
\begin{equation} f = \frac{\Gamma(.4,L_{lim},\infty)}{\Gamma(.4,0.,\infty)} \label{eq:ml}
\end{equation}
$\Gamma$ is the incomplete gamma function and $L_{lim}$ is the ratio
of $L / L^*$ represented by our estimated limiting F160W AB magnitude of
28.5.  The results are given in the first column of Table~\ref{tab3}.
The table shows that up to a redshift of 6.0 the correction is less than
50\%  and is significantly less than that for lower redshifts.  
The leveling of the error and the shallow slope of the late L$^*$ galaxy curve
in Figure ~\ref{fig4} is due to the redshift of the galaxy's high UV flux into the F160W band. 

\placetable{tab3} 

Lanzetta et al 1999 have suggested that the value of L$^*$ evolves with 
redshift z  as $(1+z)^{-1.2}$ for redshifts greater than 2.  
If we recalculate the missing luminosity with the 
evolving value of L$^*$ we get the second column in Table~\ref{tab3} which 
indicates that we are missing very substantial fractions of the luminosity at 
high redshifts. In fact by a redshift of 3.0 we are missing half of 
the luminosity.  The boundary of the objects in Figure ~\ref{fig4} is also well
fit by a template 5 L$^*$ galaxy with an extinction of E(B-V) = 0.2.  The
increasing extinction due to the decreasing rest wavelength of the F160W
band mimics an evolving L$^*$.  At this point we do not have enough information
to choose between the two effects.

There are also suggestions that the value of L$^*$ may increase with redshift,
at least to redshifts of 3 to 4.  If that is the case the fraction of missing
luminosity will be less than the first column of Table~\ref{tab3}.  In view of
the uncertainty of the true form of the luminosity function we turn to a
new method of correction described below.

\subsection{Correction via the Surface Brightness Distribution Function} \label{ssec-sbdf}

In view of the substantial uncertainties in both the aperture corrections and
luminosity function corrections described above we have developed a new technique  involving the star formation rate intensity distribution introduced 
by \cite{lanz99}.  This method calculates
a star formation rate intensity for every pixel contained in an object.
The star formation rate surface intensity x is defined as the star formation
rate in solar masses per year in a pixel divided by the proper area of the
pixel in kpc$^2$. A histogram of the distribution of the 
star formation rate surface density is then constructed by summing the proper 
areas of all pixels within a given star formation
rate and redshift interval, divided by the star formation rate interval and 
the comoving volume. From this  distribution function (denoted as h(x) by \cite{lanz99}) the star formation rate per comoving volume $\dot{\rho}$
for objects in a redshift bin is then given by

\begin{equation} \dot{\rho} = \int_{0}^{\infty} xh(x)dx\  \mathrm{
M_{\odot}\ year^{-1}\ Mpc^{-3}} \label{eq:hx}
\end{equation}

\begin{figure}
\plotone{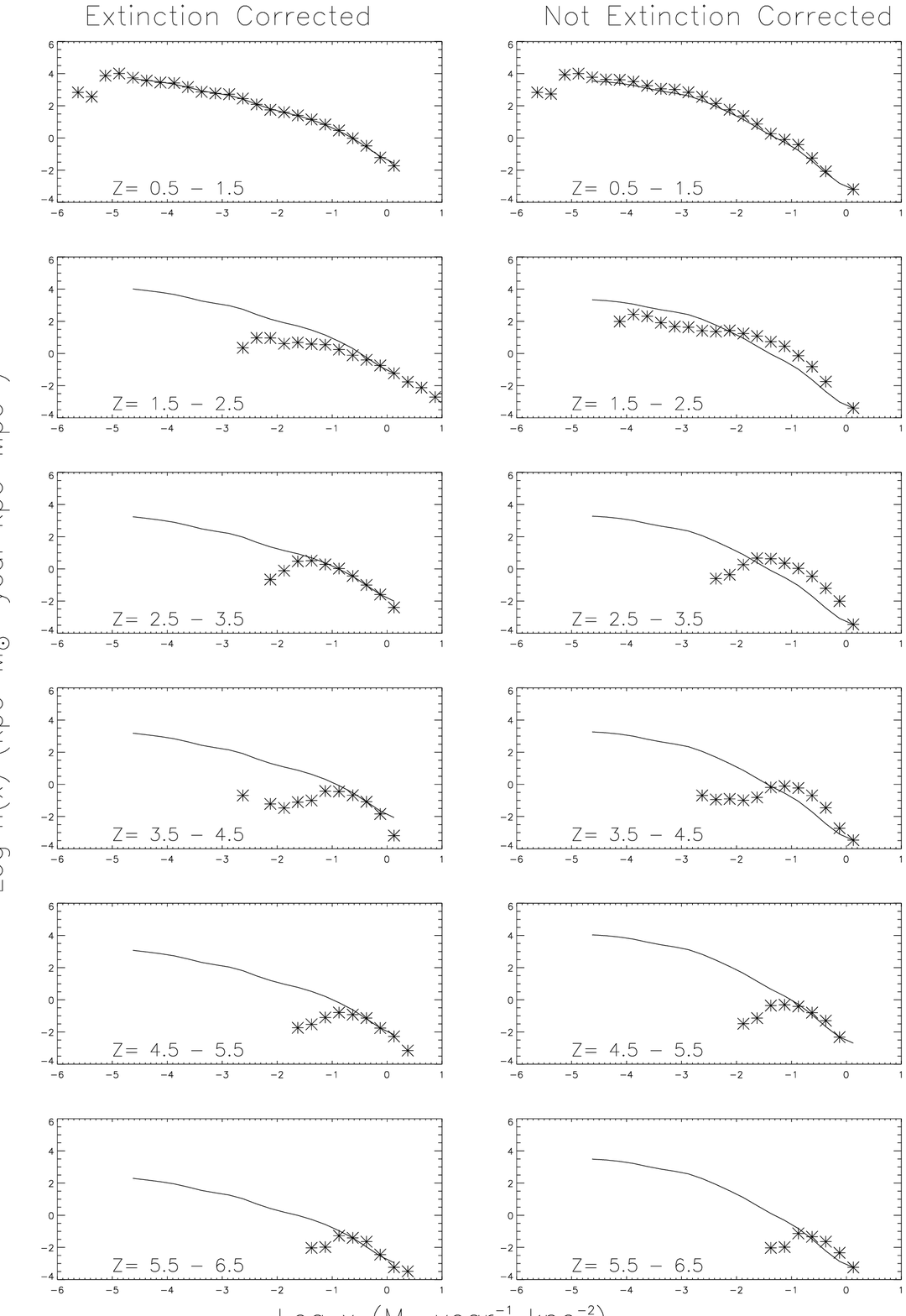}
\placefigure{fig15}
\caption{Plots of the star formation rate intensity 
distribution for the deep NICMOS HDF. The abscissa is the log of the star formation 
rate per unit area while the ordinate gives the distribution of this quantity 
per comoving volume in the redshift bin. The left hand panel is for star formation 
rates corrected for dust extinction, while for the right hand panel no 
extinction correction has been applied.} 
\label{fig15}
\end{figure}

Following \cite{lanz99} Figure~\ref{fig15} plots h(x) for those galaxies in 
our list in each of our redshift bins.  The
left hand panel uses the extinction corrected star formation value while
the right hand panel uses the uncorrected values.  In each case we have
fitted the distribution in the 0.5 to 1.5 redshift bin by a three point
smoothed curve and overplotted this curve on the other redshift bins.
The curve is adjusted by adding or 
subtracting a single value to the fitted log curve to match the bright
end of the distribution for each redshift bin. Note that the solid  curves in 
this figure are strictly  empirical matches to the data, whereas the  
solid curves in the \cite{lanz99} plot represented the distribution
expected from a bulge profile.

We next make the assumption that \emph{the shape of the extinction corrected 
h(x) distribution is the same at all redshifts and that we are successfully 
measuring the bright end of the distribution in all of the redshift bins.}
Although there is no theoretical basis for this assumption, the excellence of 
the fit in the lower redshift bins, each of which contains a completely 
independent set of objects and pixels, lends empirical support to the concept.
Note that this is not an assumption on the sizes of galaxies at various epochs.
Galaxies can be smaller (or bigger) at different epochs.  As long as their 
cumulative distributions of surface brightness of star formation are similar
the method is valid.  Also note that we are not correcting individual galaxies
but rather the total distribution of brightness of the ensemble of galaxies
in a given redshift bin.  

Under this assumption the deviation of the measured points below the empirical curve matched to the bright end of the distribution represents the missed star 
formation rate.  We then recover the true total star formation rate by
performing the integration given in equation ~\ref{eq:hx}, substituting the
value of the empirical curve whenever the measured rate dips below the
value given by the curve.  Note that the value of the integral using the
empirical curve reaches 59\% of its final value by $x = -0.25$ and
94\%  of its final value by $x = -1.25$.  We consider this new method a 
more  robust measure of the missing star formation and consider its use applied to the extinction corrected distribution as our primary method of 
estimating the missing star formation rate. The last column of Table~\ref{tab3} 
shows the incompleteness corrections from this method when the extinction
corrected distribution is used.  Note that the corrections 
for this method fall generally between the extremes of the constant L$^*$ 
corrections and the evolving L$^*$ corrections. 

The necessity of using the extinction corrected distribution rather
than the uncorrected distribution can be seen by inspection of
the right hand panel in Figure~\ref{fig15} where the  distribution
is again fitted to the curve in the 0.5 - 1.5 redshift bin.  There
is an inflection in the curve at high star formation rate intensities
that we take to represent the dimming of high star formation regions
due to extinction.  The empirical curve is then a poorer match to
the high star formation rate intensity distribution at higher redshifts
where Figure~\ref{fig5} indicates that the extinction of the galaxies
used in this study is considerably less.
It is also important to use the highest possible spatial resolution to sample
as much of the intensity distribution in each galaxy as possible.

It is important to verify the universality of our basic assumption that
the distribution function is independent of redshift. It is desirable to do 
this using the highest possible spatial resolution of both UV and Optical as
well as FIR images, in order to accurately determine the star formation
rate independently for each small pixel. For a limited set of low redshift objects this is probably currently feasible but for higher redshift objects it 
will require NGST as well as the next generation of very large ground-based 
telescopes with high performance adaptive optics working at the shortest 
feasible wavelengths.

\begin{figure}
\plotone{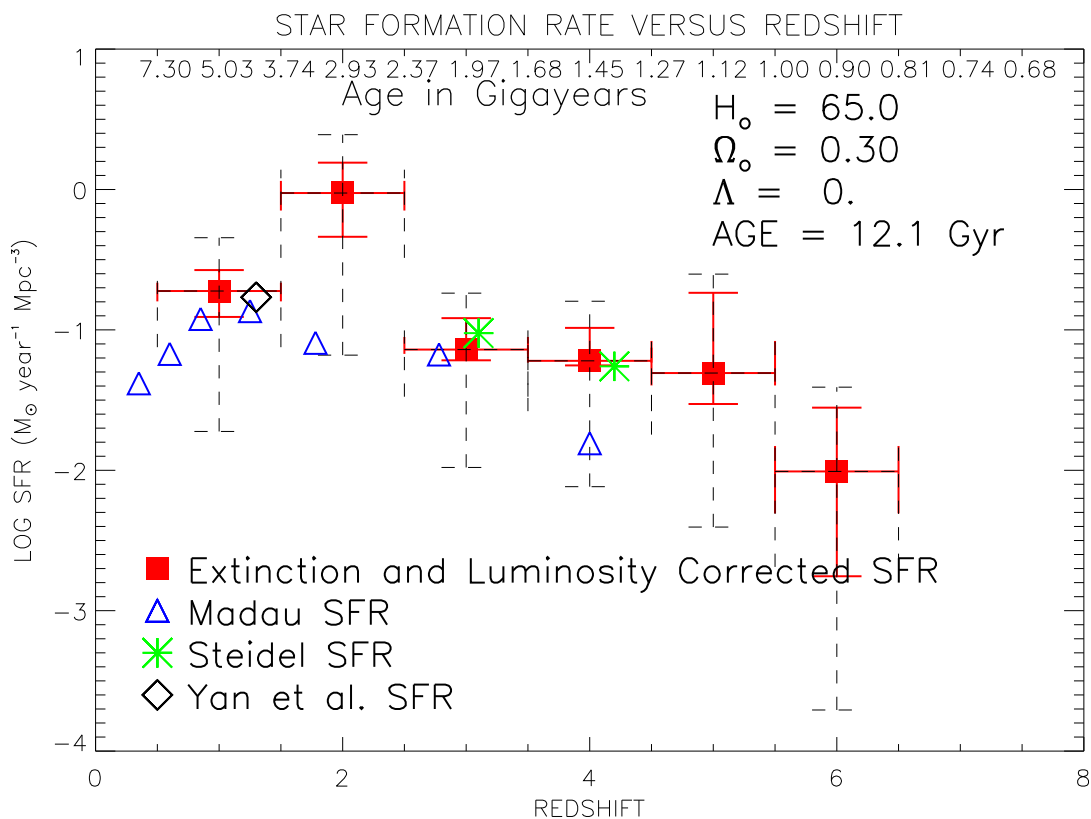}
\placefigure{fig16}
\caption{Plots of the extinction and incompleteness corrected star formation 
rate as a function of redshift.  The solid error bars indicate the photometric 
errors and the dashed error bars indicate the uncertainty in the global star
formation rate.  The star formation rate from \cite{yan99} indicated by the
diamond is for star formation in the range between a redshift of 0.8 to 2.0} 
\label{fig16}
\end{figure}

Table~\ref{tab4} gives the star formation rates derived at each level of correction for each of the methods discussed.  The column labeled Uncorrected is
the star formation rate that utilizes the zero extinction redshifts discussed 
in Section~\ref{ssec-zer}. Figure ~\ref{fig16} shows what we consider to be the correct rates based on the empirical h(x) curve adjustment of the 
extinction corrected rates.

\placetable{tab4}

\section{Sub-millimeter and Far Infrared Flux} \label{ssec-csubfarir}

A constraint on the amount of extinction we have derived for both 
individual objects and for the entire field is the amount of far infrared flux
produced.  The 450 and 850 $\micron$ flux levels in the HDF have been measured
by \cite{hug98}, and our predicted fluxes should not exceed these measured 
values, or the \cite{hug98} et al upper limits on fluxes.  In addition, 
\cite{dwek99} have determined that roughly half of the emitted flux from 
galaxies is absorbed and emitted as far infrared flux.  Although the small size 
of our field allows deviation from this general result, as discussed in 
section~\ref{ssec-uni}, we should check to see if the total far infrared flux is similar to our observed fluxes in the optical and near infrared.

Our fits to template types and dust extinction enable us to calculate
the ratio of the fluxes emitted in the optical and near infrared to that
absorbed by dust and reradiated in the far infrared and sub-millimeter 
region. The observed 850 $\micron$ flux is then given by the assumed
 temperature distribution of the dust in the source and the redshift.
Since  we do not have any observational knowledge of the actual dust temperature distribution for our galaxies, we utilize the observed Arp 220 ULIRG spectrum given in figure 4 of \cite{row93} as a standard dust spectral energy distribution.  This model has a UV optical depth of 500
which converts essentially all of the optical and UV flux into far infrared
flux.  The integrated flux of this model is scaled to the luminosity removed
by extinction for each galaxy. The model flux at the appropriate rest
wavelength is  then used to determine the expected 850 $\micron$ flux. It should be noted that variations in the actual dust SEDs from our model SED can
easily introduce errors of factor of 2 to 3 in our calculated fluxes.

Table~\ref{tab1} gives the predicted 850 $\micron$ flux in milliJanskys.
Inspection of the flux column in Table ~\ref{tab1} shows that
none of the source fluxes exceed the 2 milliJansky detection limit in the HDF 
set by \cite{hug98} although three sources, NICMOS 49.0, 166.0, and 277.211 are 
quite close.  \cite{hug98} found no sources in our observed region of the Northern HDF, consistent with the Arp 220 model.  This is not surprising since 
only five sources were detected in the Northern HDF and our area is 1/7 of the 
total area. Given the uncertainty of the fluxes we simply note that the 
predicted fluxes are consistent with the observations.

The ratio of the total power that is removed by extinction in all of the sources
to the power that is not removed by extinction is 0.8, which implies roughly 
equal amounts of power in optical-UV background and the far infrared background.
This is consistent with our current understanding of the distribution of
background power as given by \cite{dwek99}.  This result comes strictly from
our derived extinction and is not dependent on the assumed SED of the far
infrared emission. 

By summing all of the predicted 850 $\micron$ sources in our field we 
calculate a background surface brightness of $5.7 \times 10^{-10}$ watts 
m$^{-2}$ sr$^{-1}$ for $\nu I(\nu)$ at 850 $\micron$
which compares to the measured values of $5 \pm 2 \times 10^{-10}$ 
(\cite{blnb99}) and $4 \times 10^{-10}$ from COBE measurements (\cite{fix98}).
Taken together, the rough equality of the optical-UV power and FIR-power noted 
above, together with the rough equality of the predicted and observed 
submillimeter fluxes suggests to us that  i) a non-negligible fraction of 
star formation at all epochs and ii) A non-negligible portion of the 
submillimeter background is contributed by sources which are not the extremely 
super luminous sources represented by the SCUBA detections. 

A similar number at 1.6 $\micron$ can be generated from the F160W 
source fluxes which is $6 \times 10^{-9}$ watts m$^{-2}$ sr$^{-1}$ for $\nu I(\nu)$. 
This number is for detected discrete sources only.  \cite{gar96} finds 
a lower limit on resolved sources from ground based surveys at 2.2 $\micron$ 
of $7.4 \times 10^{-9}$ watts m$^{-2}$ sr$^{-1}$. Any confusion limited
surface brightness from undetected sources will be removed by our background
subtraction technique described in \cite{thm99}.

\section{Sampling errors and Overall Error Estimates}
\label{ssec-uni}

In this section we make estimates of the uncertainties
in our star formation rate determinations associated with the very small
solid angle of our sample. We consider two approaches: 1) Analytic
estimates 2) Utilization of numerical simulations.

\subsection{Analytic Estimates}
Conceptually, we distinguish between sampling errors arising from two different
sources: a) Those arising from the fact that galaxies are spatially correlated
(``large scale structure'') b) Those arising from the sampling of the
luminosity function for the small numbers of galaxies,
especially bright galaxies, in some of our bins.

\subsubsection{Large Scale Structure} 
One effect of spatial clustering is to increase the variance
in the number of objects in a cell over that expected from Poisson statistics. The fractional square root of the variance in the number of galaxies depends 
only upon the two-point correlation function and is given by \cite{peb80}.

\begin{equation}  \sigma_N/N  =   \sqrt{ 1 + N  \times I_2} / \sqrt N 
\label{eq:stat1} \end{equation}

\noindent where  

\begin{equation}  I_2 = { \int \int  \xi(r_o,\gamma) dV_1 dV_2}/V^2  
\label{eq:stat2}
\end{equation}

\noindent and where $\xi(r_o,\gamma)$ is the two point correlation function. For each of our
redshift bins it is a very good approximation to consider the volume to be a long
thin tube of square cross section with sides of (comoving) dimension D and (comoving)
length L.  Then it is straightforward to show that the expression for $I_2$ has the
form

\begin{equation}  I_2 \simeq C(\gamma) \times  (ro/D)^\gamma \times (D/L) 
\label{eq:stat3}
\end{equation}

The quantity $L/D$ is simply the number of cubes of dimension 
$D\times D\times D$ that can be placed end to end in the tube
of length L. The dimensionless coefficient $C(\gamma)$ is a double integral 
in which $dV_1$ is taken over the unit cube and $dV_2$ covers that same cube 
plus a large number of cubes on either side of this unit cube. We have evaluated this coefficient numerically which is vastly simplified by the large number of 
symmetries in this geometry. From the work of \cite{ade98} we adopt 
$\gamma$ = 1.8 and $ro = 5 h^{-1}$ Mpc. We find $C(\gamma = 1.8)$ = 8.22. 
In Rows 1, 2, 3 and 4 of Table~\ref{tab5} we give the values of N,
$1/\sqrt{N}$, I$_2$ and $\sigma _{N}/N$ computed from the above expressions for 
each of our six redshift bins. We use the value of D and L for our adopted
cosmology appropriate to the center of each bin.  For all bins except the last 
the term involving the two-point correlation dominates.

\placetable{tab5}

A difficulty with the application of equation (\ref{eq:stat1}) is that the simple
description of galaxy clustering by means of a single luminosity-independent two-point
correlation function is surely too simple. We have tried to mitigate this to some
extent by using the correlation length derived by \cite{ade98} for the most luminous
of the Ly-break galaxies. However, one can imagine an example in which each bright
galaxy carries with it a fixed number of fainter galaxies. In this case the relevant value
of N in equation (\ref{eq:stat1}) is the number of bright galaxies, not the total number
in the cell, in which case we will underestimate the fractional variance.

\subsubsection{Sampling Errors due to Small Numbers} \label{ssec-smallnum}
Even in the absence of any statistical uncertainty in the number of objects 
in a given bin, there are sampling errors. As a qualitative indication of
how important an effect this might be we show in Table~\ref{tab2} the 
percentage contribution to the total star formation rate in each bin from each 
of the most luminous objects. To obtain a quantitative estimate of the variance 
in the star formation rate we use the bootstrap resampling technique described 
in, e.g. \cite{ling86} and the references therein. The fractional variance from
this effect is shown in row 5 of Table~\ref{tab5}. We assume that the total 
uncertainty due to sampling error is obtained by combining rows 4 and 5 in 
quadrature and the resulting fractional variation due to large scale structure 
and sampling error is given in row 6.

Obviously, if the global star formation rate is dominated by very rare and 
highly luminous objects which are not present in our sample at all we cannot 
take this into account. We return to this possibility in 
section~\ref{ssec-models}.

Several groups have attempted to incorporate star formation into
N-body/hydrodynamic simulations. This approach has the virtue that all
the sampling issues discussed above can be automatically incorporated.
Ideally these simulations should cover our full redshift range, have
solid angles so large that all significant large scale structure is
averaged out, should have adequate spatial resolution to represent
objects of moderate galaxy masses, and of course, should actually model
the star formation correctly.  Such an ideal simulation would provide 
useful insights into both the role of large scale structure and the
effects of sampling errors on data sets such as ours.  Unfortunately
no simulations that we are aware of combine both adequate scale with
adequate resolution.  However existing simulations can be used for
other comparisons with our analysis, as discussed in 
section~\ref{ssec-models}.

\subsection{Overall Error Estimates} 

We have previously discussed two other sources of error in addition to
the sampling errors just discussed: (i)  Errors in the SFR resulting
from photometric errors. An inspection of Figure~\ref{fig9} shows the
16\% and 84\% confidence limits determined as described in
section~\ref{ssec-photerror} (which we take as indicative of $1\sigma$
errors) are in some cases quite asymmetric about the median and
unperturbed values.  We list these two upper and lower fractional
errors in rows 7 and 8 of table~\ref{tab5}. ii) Errors resulting from
the high degree of degeneracy between stellar population type and
internal reddening. These are over and above any errors in photometry
and would arise from the failure of individual galaxies to be
adequately represented by our set of 51 templates. Based upon the
discussion of section~\ref{ssec-tgen} we estimated the fractional SFR
error to be $\pm$ a factor of 2 in the SFR. For completeness we list
these values in rows 9 and 10 of table~\ref{tab5}. We do not include
any error estimates for the accuracy of the incompleteness corrections.
These corrections, based upon the histograms of Figure~\ref{fig14} are
on the order of a factor of two at high redshift, as shown by
table~\ref{tab4}. The error in the correction should be significantly
less than the correction which means it will be dominated by the
sampling errors. 

To get the total upper fractional error in the star formation rate we simply 
add the upper fractional errors in Table~\ref{tab4} in quadrature.  If we also
do this for the lower error the procedure leads to a negative star formation 
rate in some bins which is not physical.  To compute the lower bound we
simply multiplied the fractional lower errors in Table~\ref{tab4} 
together.  This may be somewhat pessimistic since it assumes that all lower 
error fractions contribute their full weight to the fractional error.  The
calculated upper and lower errors are given in the last two rows of
Table~\ref{tab4}.  These errors, when translated into SFR values, are 
plotted by the dashed error bars in Figure~\ref{fig16}.

\section{Discussion} \label{sec-disc}

In this section we compare our results with other studies and look at a few
of the implications of our results.  We do note that our measured star
formation rate at a redshift of 5 is equal to that at redshift 3 thus 
satisfying the criterion for reionization by stars stated in \cite{mad99} and discussed in the introduction.  Of course this would require that a significant 
fraction of the Lyman continuum flux actually escape from these star forming 
galaxies, and it is exceedingly difficult to know if this is the case. As noted 
earlier, in our templates we have assumed that no Lyman continuum photons 
escape any of the galaxies.

\subsection{Comparison with Other Optically-Derived Results} \label{ssec-others}

Except for a large discrepancy near a redshift of 2 our star formation rates in 
Figure~\ref{fig16} are roughly consistent with the rates found by 
\cite{mad99}. Our redshift 1 results are also consistent with H$\alpha$ derived
rates of \cite{yan99} that cover the redshift range between 0.8 and 2.0 but 
our rate for a redshift of 2 is significantly higher than this due, as noted
previously, to two very luminous reddened galaxies. The rates from a redshift of 3 through 5 are consistent with a constant star formation rate and agree with the values found by \cite{stei99} based on data from a much larger area. 
All of the rates not
based on our data have been adjusted to our cosmology. The solid line error bars
on our results come from the techniques discussed in section~\ref{sec-err}.  As 
such they represent the errors associated with the star formation rate 
\emph{in the NICMOS observed region of the Northern Hubble Deep Field}.  The
dashed line error bars represent the errors in extrapolating this result
to the universe in general taking account of other sources of error, including sampling errors (as discussed in section~\ref{ssec-uni}) in addition to the photometric error bars plotted in figure~\ref{fig6}. 

\subsection{Comments on Other Star Formation History Work} \label{ssec-models}

We begin with a brief philosophical comment: As mentioned at the
conclusion of Section~\ref{ssec-smallnum}, it is certainly conceivable that 
the average comoving density of star formation in any or all redshift bins
is completely dominated by  objects totally missing from our
sample, either because they are very rare, and thus missing from
our small solid angle survey, and/or because they are so heavily
obscured that they are too faint to be seen even in our deep
F110W and F160W images. Indeed, a current lively debate centers
on just this latter possibility, as discussed further in
Section~\ref{ssec-scubas}. 

While these possibilities and the global comoving star formation
rate are obviously important issues, they are hardly the only issues.
A complete picture of the history of star formation should,
at the very least, encompass delineation of, and explanation for,
a much more complex ``star formation distribution function'', e.g.
$\phi(SFR,fIR,z)$ giving the number of galaxies at redshift
z undergoing star formation at a rate SFR with a fraction fIR
of the luminosity being converted to the far infrared. Thus, 
despite our limited solid angle and the fact that the objects
in our survey are near IR--selected we believe such surveys
as this are of interest because they probe to low values of the
SFR for objects which are not heavily obscured. We believe the
deep NICMOS survey to be 50\% complete at $m_{AB}F160W$ of 28.5
through an 0.6$\arcsec$ aperture. At a redshift of 3
a  template fit of 5.5 and E(B-V) of 0.06, typical for our sample,
the inferred star formation rate is 0.4 $M_\odot/yr$.

\subsubsection{The Theoretical Model of Weinberg et al}

\cite{wein99} present SPH numerical simulations to which they adjoin a
recipe for star formation and energy deposition by supernovae.\footnote
{We are very indebted to David Weinberg for providing details of these
simulations and for permission from Dr. Weinberg as well as N. Katz, L.
Hernquist and R. Dave to quote these results.} By
combining models with different parameters they  extend the resolution
down to baryon masses of order $10^8$ (see their Table 1) and to star
formation rates as low as 0.1 solar masses per year (see their Figure
1), though, as the authors point out, the flattening of the
distribution at these low values may indicate they are
resolution--limited at somewhat higher values.

We attempt a comparison between these models and our results as
follows: (1) We restrict analysis to the objects for which we assigned
redshifts in the range 2.5 to 4.5. In this range we are generally
directly observing sufficiently far into the rest UV that
we are less at the mercy of the template fitting than at lower
redshifts, while at higher redshifts our sample is too small.
(2) We omit objects  for which the template fit is less than 4.0. 
In these cases we are looking at older 
populations where the bulk of the star formation has already
taken place and where the SPH models may not apply. 
(3) The SFR is at least 0.5 at which point we may consider our
sample to be fairly complete as suggested by the example above.
We find 57  objects in Table~\ref{tab1}
satisfying these criteria.  We normalize the cumulative
distribution of the number of objects as a function of their star formation 
rate for both our sample and the \cite{wein99} models.  As noted above, for typical template and extinction values
of the objects in this redshift range, we should be about 50\% complete
at a star formation rate of 0.4 $M_\odot/yr$. Thus, we normalize cumulative
distributions defining them to be zero at the slightly higher value
of 0.5 $M_\odot/yr$ and 1.0 at a star formation rate of $\infty$.

We may then compare the {\it shapes} of these distributions by examining
the ratio of the star formation rates at the 25th percentile and 50th
percentiles, and similarly at the 25th and 75th percentiles. We may
also compare the absolute value of the star formation rate at the 25th
percentile. Table~\ref{tab6} presents the result of these comparisons.
Given the small number statistics of our sample and the errors
discussed earlier this is rather remarkable agreement.

\placetable{tab6}

A difference does arise when we compare our actual
observed surface density from this sample with the simulations.
In our survey we find a surface density of 60 objects
per square arc minute over this range. Adding together
the number of objects in the redshift bins of 3 and 4 down
to a SFR = 0.5 from \cite{wein99} their simulations predict
a surface density of 400 objects per square arc minute, about
seven times  higher than we observe. Since their
distribution is flattening at this point due to possible
resolution effects, the discrepancy might be higher still.

It is not clear that this is a significant discrepancy.  Our
incompleteness corrections described in Section~\ref{sec-csfrh}
correct for missed star formation not for the number of objects
contributing to that rate. A small increase of 0.1 over our typical
value of E(B-V) of 0.06 will result in more than a half magnitude
of dimming at the rest wavelength of the F160W band in the redshift 
range from 3 to 4. Inspection of Figure~\ref{fig4} shows that this would 
remove a large number of faint sources from our view.  Since
the highest number of sources reside near the boundary of
our detection limit it is difficult for the observations to
accurately determine the absolute number.

Since the actual star formation rate is dominated by the brightest
sources and the number statistics by the faintest sources, it is  
appropriate to compare our observations with the calculated star 
formation rates given in Figure 2 of \cite{wein99}.  The log of the 
star formation rate at redshifts of 3 to 4 in that figure is -0.5
while our value is a -1.1.  Our ``1 $\sigma$'' error bar extends to -0.7
and of course the log of our star formation rate at a z of 2 is
0.0.  There is a general trend of the numerical simulations to
have somewhat higher star formation rates than we have observed 
but within the probable errors of both the simulations and the
observation the gap is not wide.

\subsubsection{SCUBA sources: Ly Break galaxies or an Optically Hidden Population} \label{ssec-scubas}

As indicated above, a lively debate is occurring between those who advocate
 that the bulk of the star formation at high redshift is occurring in luminous
heavily obscured objects which can only be detected at submillimeter wavelengths
and those who argue that optically--selected galaxies, even though subject
to significant extinction, can account for both the bulk of the global star 
formation and indeed the bulk of the submillimeter flux as well. Our analysis
in Section~\ref{ssec-csubfarir} favors the latter point of view. 
After that analysis had been
completed we became aware of the work of \cite{ade2000} who strongly favor
this latter point of view, using a completely different sample and line of argument.

\cite{ade2000} assemble data on narrow band fluxes and
indices over a wide range of
the electromagnetic spectrum from local sources to reconstruct typical SEDs of
star forming galaxies and infer the shape of the dust emission and total star formation
rate. They examine the correlations between these various pieces of data and
examine the much more limited evidence at higher redshifts to see if these correlations
hold. While the evidence is not overwhelming, a plausible case is made that they
do, and in particular that the slope of the UV spectral index, $\beta$, is a good
indicator of the submillimeter flux and along with the UV flux, can be used estimate
the star formation rate. 

As we have noted in Section~\ref{sec-err}, several combinations of
extinctions and templates (and thus the total star formation rate and the fraction
of energy converted to the far IR) produce very similar optical spectral energy 
distributions.
Thus, as these authors themselves state, it is not clear why such a correlation should hold
involving such complex situations as variable star formation histories, amount and
geometry of dust obscuration etc.  It is possible that the degeneracy which we explored in Section~\ref{sec-err} 
could be broken as and when sufficiently large light gathering power is 
available to examine spectral features in the rest UV.
 Nevertheless, our estimates of the star formation
rates in these high redshift bins agree well with those of \cite{stei99}, and the
two independent conclusions on this issue seem to us to favor the point of view
that observations in the optical and near infrared can determine the star 
formation rates accurately. With NICMOS we also easily observe sources which 
have more than $95\%$ of their luminosity being reemitted in the far infrared
and whose 850 {\micron} flux levels are below the SCUBA detection limit.
Whether the rare, strong SCUBA sources are produced by objects which we could
not detect in our F160W and F110W images will depend simply on whether the extinction to these sources is significantly greater than the amount needed
to make most of their luminosity appear at far infrared wavelengths.

\section{Conclusions}   \label{sec-conc}

An important conclusion from our work is the relatively flat star formation rate between a redshift of 3 and 5 when corrections for both extinction and missed 
flux are taken into account.  This implies that star formation occurs at a 
reasonably steady rate in the early universe. However it is still 
true that the vast majority of stars formed at redshifts less than 2 due to the 
much greater amount of time available at low redshift values.   Explaining
this behavior can provide a constraint to models of star and galaxy formation.

The ability of our observed sources to reasonably reproduce the observations
of the net background at $850 \micron$ indicates that the type of sources 
found in our study contribute significantly to the global star formation rate.
Increases in the star formation rates by a factor of 10 due to extremely 
luminous objects represented by the SCUBA detections appears unlikely.

Our high star formation rate at z = 2 is due to the presence of two 
possible ULIRGs in our field which contribute 80 \% of the measured 
star formation rate.  The presence of two ULIRGs in our small field is highly 
unlikely at the local density of ULIRGs and it may imply a greatly increased 
number of merger events leading to high luminosity at a redshift of 2.

\acknowledgments
This work is supported in part by NASA grant NAG 5-3042.  This work
utilized observations with the NASA/ESA Hubble Space Telescope,
obtained at the Space Telescope Science Institute, which is operated by
the Association of Universities for Research in Astronomy under NASA
contract NAS5-26555. We wish to thank an anonymous referee for
suggesting the artificial data test for the partial template extinction
degeneracy and for comments that have improved the quality of this
work. Two of us (RIT and RJW) wish to acknowledge extremely helpful
discussions with David Koo, Alex Szalay, Daniela Calzetti and David
Weinberg.  RIT wishes to acknowledge several productive discussions
with Ken Lanzetta, Mathias Steinmetz, Mark Dickinson, Neta Bahcall, and
Piero Madau. RJW wishes to acknowledge helpful discussions with Ray
Carlberg,  David Elbaz and Mauro Giavalisco.

\clearpage

\begin{deluxetable}{cccccccccccccc}
\tablecaption{Listing of measured quantities \label{tab1}}
\tablewidth{0pt}
\scriptsize
\tablehead{ 
\colhead{NICMOS} & \colhead{WFPC} & \colhead{z} & \colhead{E(B-V)} & \colhead{SFR} &  \colhead{Lum.} & \colhead{frac.\tablenotemark{a}}  & \colhead{SCUBA}  &  \colhead{T \tablenotemark{b}} & \colhead{$\chi^2$} & \colhead{Tot.\tablenotemark{c}} & \colhead{Ap.\tablenotemark{d}} & \colhead{RA} & \colhead{DEC} \\ \colhead{ID} & \colhead{ID} & \colhead{} & \colhead{} & \colhead{M$_\odot$} & \colhead{L$_\odot$} & \colhead{} & \colhead{850 \micron} & \colhead{} & \colhead{ }& \colhead{mag} & \colhead{mag} & \colhead{12\fh} & \colhead{+62\fdg} \\ \colhead{} & \colhead{} & \colhead{} & \colhead{} & \colhead{yr$^{-1}$} & \colhead{} & \colhead{} & \colhead{flux} & \colhead{} & \colhead{} & \colhead{} & \colhead{} & \colhead{36\fm} \\ \colhead{} & \colhead{} & \colhead{} & \colhead{} & \colhead{} & \colhead{} & \colhead{} & \colhead{mJy} }
\startdata
 1.00000 &\nodata& 1.52& 0.00& 0.356 &7.06E+08& 0.00 &0.00& 5.6& 1.0  & 27.1  & 27.8 &40.81&12:9.60 \nl
 2.00000 &4-830.0& 3.84& 0.10& 2.710 &7.53E+09& 0.46 &3.96E-03& 5.5& 2.0  & 27.0  & 27.2 &40.96&12:10.8 \nl
 502.000 &4-807.0& 3.84& 0.00& 1.146 &3.21E+09& 0.00 &0.00& 5.6& 1.7  & 27.6  & 27.8 &41.13&12:12.2 \nl
 5.00000 &4-822.0& 0.16& 0.70& 0.615 &1.28E+09& 0.95 &2.82E-02& 6.0& 2.6  & 24.7  & 25.2 &41.14&12:10.6 \nl
 10.0000 &4-813.2& 3.60& 0.00& 0.897 &4.68E+09& 0.00 &0.00& 4.8& 0.6  & 26.8  & 26.9 &41.42&12:6.80 \nl
 11.0000 &4-790.0& 3.60& 0.00& 1.137 &2.71E+09& 0.00 &0.00& 6.0& 0.2  & 27.7  & 27.9 &41.43&12:11.4 \nl
 14.0000 &4-767.0& 0.72& 0.02& 0.019 &8.48E+09& 0.02 &6.26E-04& 1.1& 0.8  & 21.7  & 22.0 &41.48&12:15.0 \nl
 15.0000 &4-813.0& 3.36& 0.00& 2.917 &8.97E+09& 0.00 &0.00& 5.3& 1.2  & 25.9  & 26.2 &41.53&12:6.80 \nl
 16.0000 &4-794.0& 0.80& 0.10& 0.432 &1.15E+09& 0.47 &1.74E-03& 5.6& 0.4  & 26.1  & 26.5 &41.55&12:8.10 \nl
 1085.00 &\nodata& 0.48& 0.20& -0.00 &1.51E+08& 0.60 &\nodata& 6.0& 6.8&\nodata&\nodata &41.60&12:6.20 \nl
 18.0000 &\nodata& 0.64& 0.50& 0.464 &1.30E+09& 0.92 &2.60E-03& 6.0& 0.9  & 27.5  & 27.7 &41.62&12:12.8 \nl
 21.0000 &4-739.0& 0.32& 0.50& 0.243 &5.42E+08& 0.92 &2.73E-03& 6.0& 0.3  & 26.9  & 27.3 &41.64&12:16.1 \nl
 1084.00 &\nodata& 3.36& 0.00& 3.595 &1.34E+09& 0.00 &0.00& 6.0& 11.  & 28.4  & 30.7 &41.66&12:6.40 \nl
 26.1200&4-795.11& 0.48& 0.30& 1.443 &2.92E+09& 0.77 &1.24E-02& 5.9& 1.7  & 25.1  & 25.6 &41.86&12:7.00 \nl
 26.1000 &4-795.0& 0.40& 0.10& 0.460 &7.12E+09& 0.23 &9.91E-03& 3.0& 2.5  & 20.8  & 21.4 &41.94&12:5.40 \nl
 1081.00 &4-748.0& 4.64& 0.02& 0.782 &2.44E+09& 0.12 &3.42E-04& 5.5& 0.5  & 28.3  & 28.4 &41.96&12:9.10 \nl
 1086.00 &\nodata& 0.48& 0.20& 0.107 &1.68E+08& 0.56 &6.98E-04& 5.8& 1.9  & 27.4  & 28.3 &42.05&12:3.10 \nl
 514.000 &4-711.2& 0.96& 0.40& 0.522 &1.57E+09& 0.88 &2.34E-03& 6.0& 0.8  & 28.1  & 28.1 &42.07&12:14.8 \nl
 515.000 &4-709.2& 2.48& 0.00& 0.246 &6.87E+08& 0.00 &0.00& 5.5& 2.5  & 27.9  & 28.2 &42.15&12:13.8 \nl
 27.0000 &4-665.0& 1.44& 0.20& 3.602 &9.16E+09& 0.60 &8.95E-03& 6.0& 1.2  & 25.9  & 26.1 &42.17&12:23.0 \nl
 29.0000 &4-769.0& 0.96& 0.20& 1.362 &3.07E+09& 0.55 &4.19E-03& 5.7& 1.7  & 25.7  & 26.2 &42.29&12:1.40 \nl
 31.0000 &4-671.0& 0.96& 0.30& 1.556 &3.85E+09& 0.75 &6.34E-03& 5.8& 0.9  & 26.0  & 26.4 &42.36&12:19.0 \nl
 32.0000 &4-690.0& 1.12& 0.40& 2.459 &5.39E+09& 0.88 &8.05E-03& 6.0& 1.5  & 26.5  & 26.9 &42.37&12:13.9 \nl
 33.0000 &4-743.0& 0.48& 0.10& 0.044 &1.71E+08& 0.50 &2.58E-04& 5.8& 5.6  & 28.2  & 28.1 &42.47&12:1.20 \nl
 34.0000 &4-715.0& 4.00& 0.08& 1.626 &4.44E+09& 0.35 &2.06E-03& 5.2& 0.3  & 27.4  & 27.8 &42.54&12:5.60 \nl
 35.0000 &4-687.0& 0.08& 0.50& 0.018 &4.88E+07& 0.92 &3.23E-03& 6.0& 8.3  & 26.9  & 27.1 &42.58&12:11.4 \nl
 36.0000 &4-619.0& 2.96& 0.10& 1.219 &4.44E+09& 0.42 &2.07E-03& 5.3& 0.4  & 26.9  & 27.0 &42.61&12:25.3 \nl
 1067.00 &\nodata& 3.92& 0.00& 1.417 &1.68E+09& 0.00 &0.00& 5.5& 0.7  & 27.5  & 28.6 &42.61&12:17.0 \nl
 37.0000 &4-725.0& 1.84& 0.00& 0.661 &2.16E+09& 0.00 &0.00& 4.9& 0.9  & 25.4  & 26.1 &42.64&12:4.10 \nl
 38.0000 &4-636.0& 0.64& 0.40& 1.675 &4.15E+09& 0.88 &8.92E-03& 6.0& 1.5  & 25.9  & 26.2 &42.65&12:20.9 \nl
 43.0000 &4-707.0& 0.80& 0.08& 0.404 &8.07E+08& 0.43 &1.38E-03& 5.8& 2.3  & 26.7  & 27.3 &42.77&12:3.90 \nl
 528.000 &4-616.0& 3.76& 0.10& 1.825 &3.16E+09& 0.47 &2.74E-03& 5.6& 2.6  & 27.3  & 28.0 &42.79&12:22.7 \nl
 1071.00 &4-655.0& 2.72& 0.00& 0.422 &9.99E+08& 0.00 &0.00& 6.0& 4.8  & 29.1  & 29.3 &42.86&12:13.6 \nl
 47.1000 &4-581.0& 1.92& 0.10& 6.356 &1.51E+10& 0.50 &1.33E-02& 5.8& 2.3  & 25.0  & 25.3 &42.87&12:27.8 \nl
 45.0000 &4-716.0& 1.60& 0.00& 0.075 &3.65E+08& 0.00 &0.00& 5.4& 0.6  & 28.6  & 28.4 &42.88&12:0.00 \nl
 46.0000 &4-697.0& 2.48& 0.02& 1.818 &5.39E+09& 0.12 &1.16E-03& 5.4& 4.1  & 25.8  & 26.0 &42.90&12:3.50 \nl
 48.0000 &4-606.0& 2.80& 0.20& 2.997 &5.49E+09& 0.60 &5.85E-03& 6.0& 0.1  & 27.2  & 27.7 &42.91&12:22.8 \nl
 49.0000 &4-656.0& 0.56& 0.50& 67.57 &1.50E+11& 0.91 &4.99E-01& 5.9& 2.4  & 21.0  & 21.8 &42.91&12:16.3 \nl
 47.2000 &4-581.2& 1.84& 0.02& 1.502 &4.31E+09& 0.11 &1.00E-03& 5.3& 1.8  & 25.4  & 25.8 &42.96&12:26.6 \nl
 51.0000 &4-660.0& 2.32& 0.02& 3.257 &7.65E+09& 0.13 &1.28E-03& 5.6& 3.5  & 25.3  & 25.7 &42.97&12:11.5 \nl
 1073.00 &4-649.0& 1.12& 0.02& 0.094 &2.29E+08& 0.14 &5.23E-05& 5.9& 2.9  & 28.7  & 29.1 &43.09&12:11.2 \nl
 54.0000 &4-664.0& 1.84& 0.00& 0.365 &8.80E+08& 0.00 &0.00& 5.5& 3.6  & 27.1  & 27.6 &43.11&12:7.20 \nl
 56.0000 &4-537.0& 2.40& 0.00& 0.258 &7.49E+08& 0.00 &0.00& 5.2& 8.4  & 27.2  & 27.7 &43.12&12:31.0 \nl
 57.0000 &4-554.0& 0.48& 0.20& 1.049 &2.92E+09& 0.53 &6.92E-03& 5.6& 3.0  & 24.4  & 24.7 &43.13&12:28.1 \nl
 59.0000 &4-599.0& 2.80& 0.00& 2.622 &7.78E+09& 0.00 &0.00& 5.4& 3.6  & 25.7  & 25.9 &43.19&12:19.2 \nl
 1062.00 &4-600.0& 4.48& 0.00& 1.725 &2.46E+09& 0.00 &0.00& 6.0& 7.6  & 27.3  & 28.0 &43.25&12:17.7 \nl
 61.0000 &4-590.0& 2.08& 0.30& 21.32 &4.89E+10& 0.79 &5.58E-02& 6.0& 1.1  & 24.8  & 25.1 &43.30&12:18.3 \nl
 1064.00 &4-591.0& 0.56& 0.30& 0.151 &3.09E+08& 0.77 &9.51E-04& 5.9& 1.2  & 27.7  & 28.2 &43.34&12:17.2 \nl
 63.0000 &4-605.0& 2.80& 0.00& 0.629 &2.41E+09& 0.00 &0.00& 5.7& 1.6  & 27.7  & 27.5 &43.35&12:15.9 \nl
 64.0000 &4-677.0& 0.32& 0.20& 0.080 &1.69E+08& 0.58 &5.85E-04& 5.9& 2.5  & 27.2  & 27.8 &43.40&12:1.00 \nl
 1063.00 &4-583.0& 0.48& 0.30& 0.155 &2.60E+08& 0.79 &1.32E-03& 6.0& 1.9  & 28.5  & 29.2 &43.46&12:17.0 \nl
 68.0000 &4-565.0& 0.56& 0.40& 21.52 &4.71E+10& 0.86 &1.51E-01& 5.9& 1.4  & 22.5  & 23.0 &43.62&12:18.3 \nl
 70.0000 &4-669.0& 0.48& 0.30& 0.116 &3.74E+08& 0.79 &9.97E-04& 6.0& 0.9  & 28.3  & 28.4 &43.69&11:57.0 \nl
 71.0000 &4-572.0& 0.48& 0.30& 1.051 &1.97E+09& 0.75 &9.08E-03& 5.8& 1.1  & 25.1  & 25.8 &43.70&12:15.6 \nl
 72.0000 &4-598.0& 0.56& 0.30& 0.268 &6.46E+08& 0.75 &1.70E-03& 5.8& 2.0  & 26.7  & 27.1 &43.71&12:11.0 \nl
 74.0000 &4-513.0& 1.04& 0.20& 0.266 &6.82E+08& 0.55 &6.98E-04& 5.7& 1.3  & 27.7  & 28.0 &43.75&12:25.2 \nl
 554.000 &4-492.0& 4.16& 0.02& 0.362 &2.24E+09& 0.15 &1.21E-04& 6.0& 0.3  & 29.4  & 28.6 &43.79&12:29.1 \nl
 75.0000 &4-479.0& 1.12& 0.06& 1.512 &3.70E+09& 0.35 &2.12E-03& 5.8& 0.4  & 25.7  & 26.1 &43.79&12:32.1 \nl
 76.0000 &4-525.0& 0.56& 0.10& 0.141 &6.85E+08& 0.54 &6.02E-04& 6.0& 1.9  & 28.0  & 27.6 &43.81&12:22.4 \nl
 77.0000 &4-486.0& 0.48& 0.50& 0.232 &6.32E+08& 0.92 &2.32E-03& 6.0& 0.5  & 27.8  & 28.0 &43.83&12:29.7 \nl
 1072.00 &4-595.0& 2.40& 0.06& 0.426 &7.14E+08& 0.33 &7.27E-04& 5.6& 1.6  & 27.8  & 28.6 &43.86&12:8.00 \nl
 562.000 &4-663.0& 4.88& 0.06& 0.925 &2.27E+09& 0.37 &9.16E-04& 6.0& 3.1  & 28.8  & 28.9 &43.91&11:54.4 \nl
 79.0000 &4-626.0& 1.84& 0.00& 0.423 &9.26E+08& 0.00 &0.00& 5.8& 0.6  & 27.6  & 28.0 &43.91&12:2.20 \nl
 81.1000 &4-576.0& 3.36& 0.00& 5.878 &1.39E+10& 0.00 &0.00& 5.5& 1.1  & 25.4  & 25.8 &44.02&12:9.50 \nl
 80.0000 &4-444.0& 0.72& 0.00& 0.015 &9.59E+07& 0.00 &0.00& 4.2& 0.8  & 27.1  & 27.5 &44.02&12:36.5 \nl
 1080.00 &4-612.0& 0.00& 0.00& 0.000 &1.47E+02& 0.00 &0.00& 4.4& 0.4  & 27.8  & 28.4 &44.03&12:3.00 \nl
 83.0000 &\nodata& 1.12& 0.00& 0.074 &2.84E+08& 0.00 &0.00& 5.4& 2.3  & 27.8  & 27.9 &44.05&12:23.1 \nl
 81.2000 &4-576.2& 1.52& 0.20& 2.869 &6.61E+09& 0.58 &6.38E-03& 5.9& 0.5  & 26.0  & 26.3 &44.07&12:9.20 \nl
 86.0000 &4-500.0& 1.28& 0.00& 0.055 &2.90E+08& 0.00 &0.00& 5.2& 1.4  & 27.9  & 27.8 &44.08&12:23.4 \nl
 87.0000 &4-653.0& 2.48& 0.00& 0.553 &1.30E+09& 0.00 &0.00& 5.6& 4.4  & 27.2  & 27.6 &44.10&11:53.3 \nl
 1044.00 &4-500.0& 2.32& 0.02& 0.723 &2.09E+09& 0.15 &2.85E-04& 6.0& 3.6  & 27.6  & 27.6 &44.12&12:23.1 \nl
 88.0000 &4-430.0& 0.72& 0.06& 0.870 &2.58E+09& 0.26 &3.81E-03& 5.0& 3.5  & 23.5  & 24.3 &44.19&12:40.4 \nl
 1041.00 &4-485.0& 0.48& 0.02& -0.00 &8.56E+07& 0.15 &\nodata& 6.0& 1.3&\nodata&\nodata &44.23&12:23.1 \nl
 568.000 &4-577.0& 3.92& 0.10& 1.160 &2.93E+09& 0.46 &1.62E-03& 5.5& 0.3  & 28.1  & 28.4 &44.26&12:4.60 \nl
 1079.00 &4-589.0& 0.16& 0.40& 0.022 &3.89E+07& 0.88 &9.39E-04& 6.0& 1.7  & 28.1  & 28.8 &44.31&12:1.90 \nl
 90.0000 &4-580.0& 1.04& 0.30& 0.603 &1.26E+09& 0.75 &2.09E-03& 5.8& 0.7  & 27.2  & 27.7 &44.33&12:3.20 \nl
 91.0000 &4-524.0& 1.84& 0.30& 2.455 &4.27E+09& 0.79 &8.27E-03& 6.0& 0.4  & 26.9  & 27.6 &44.35&12:14.0 \nl
 92.0000 &4-499.0& 4.88& 0.20& 5.781 &1.37E+10& 0.53 &9.26E-03& 5.6& 7.9  & 27.4  & 27.7 &44.44&12:17.2 \nl
 571.000 &4-534.0& 0.96& 0.40& 0.658 &1.29E+09& 0.88 &2.94E-03& 6.0& 2.5  & 27.9  & 28.5 &44.46&12:9.80 \nl
 94.0000 &4-627.0& 0.16& 0.60& 0.170 &3.50E+08& 0.95 &7.82E-03& 6.0& 1.2  & 26.1  & 26.7 &44.47&11:53.3 \nl
 95.0000 &4-438.0& 1.12& 0.06& 0.327 &9.53E+08& 0.28 &5.10E-04& 5.2& 0.0  & 26.0  & 26.5 &44.49&12:30.5 \nl
 96.0000 &\nodata& 4.88& 0.20& 6.143 &1.15E+10& 0.55 &9.74E-03& 5.7& 1.7  & 27.4  & 27.9 &44.54&12:36.1 \nl
 97.0000 &\nodata& 1.28& 0.60& 18.83 &3.57E+10& 0.87 &1.05E-01& 5.5& 3.2  & 23.6  & 24.3 &44.56&12:15.4 \nl
 98.0000 &4-527.0& 0.40& 0.50& 0.560 &1.43E+09& 0.92 &8.03E-03& 6.0& 2.5  & 26.5  & 26.8 &44.57&12:9.60 \nl
 100.000 &4-472.0& 2.96& 0.00& 0.875 &2.60E+09& 0.00 &0.00& 5.2& 1.2  & 26.7  & 27.1 &44.59&12:20.4 \nl
 579.000 &4-546.0& 1.28& 0.20& 0.447 &1.03E+09& 0.60 &1.41E-03& 6.0& 0.0  & 28.0  & 28.4 &44.62&12:4.20 \nl
 101.000 &4-632.0& 0.24& 0.06& 0.000 &8.47E+06& 0.27 &0.00& 5.1& 0.5&\nodata  & 29.0 &44.62&11:48.8 \nl
 1059.00 &\nodata& 1.68& 0.30& 1.232 &1.91E+09& 0.77 &2.96E-03& 5.9& 2.7  & 27.5  & 28.3 &44.63&12:14.0 \nl
 102.000 &4-445.0& 1.84& 0.30& 80.45 &1.74E+11& 0.75 &2.74E-01& 5.8& 3.6  & 22.8  & 23.3 &44.64&12:27.4 \nl
 103.000 &4-625.0& 4.48& 0.10& 21.19 &5.47E+10& 0.39 &4.28E-02& 5.1& 0.5  & 24.8  & 25.4 &44.66&11:50.5 \nl
 1046.00 &4-472.2& 0.00& 0.50& 0.000 &2.55E+03& 0.91 &0.00& 5.9& 0.7  & 27.4  & 28.2 &44.67&12:20.4 \nl
 584.000 &\nodata& 0.32& 0.40& 0.009 &3.12E+07& 0.69 &1.42E-04& 5.0& 1.8  & 27.5  & 28.1 &44.68&12:14.2 \nl
 586.000 &\nodata& 5.68& 0.06& 2.998 &5.02E+09& 0.37 &2.18E-03& 6.0& 0.0  & 27.5  & 28.0 &44.71&12:20.0 \nl
 107.000 &\nodata& 5.92& 0.00& 2.391 &5.29E+09& 0.00 &0.00& 5.2& 0.0  & 27.1  & 27.7 &44.72&12:18.8 \nl
 109.000 &4-579.0& 0.48& 0.20& 1.103 &3.00E+09& 0.56 &7.15E-03& 5.8& 0.8  & 24.9  & 25.2 &44.74&11:57.1 \nl
 110.000 &4-603.0& 2.56& 0.02& 2.072 &4.88E+09& 0.12 &1.24E-03& 5.4& 1.9  & 25.7  & 26.2 &44.74&11:54.4 \nl
 1031.00 &4-422.0& 1.92& 0.00& 0.117 &3.85E+08& 0.00 &0.00& 5.4& 2.0  & 28.1  & 28.3 &44.77&12:29.3 \nl
 588.000 &4-483.0& 0.96& 0.08& 0.102 &2.84E+08& 0.36 &2.62E-04& 5.3& 0.2  & 27.3  & 27.8 &44.79&12:14.5 \nl
 114.000 &4-558.0& 0.48& 0.40& 5.404 &1.01E+10& 0.84 &5.23E-02& 5.8& 4.3  & 23.3  & 24.0 &44.83&12:0.20 \nl
 116.000 &4-509.0& 1.12& 0.00& 0.020 &3.52E+08& 0.00 &0.00& 2.9& 2.1  & 26.5  & 26.7 &44.84&12:7.60 \nl
 901.000 &\nodata& 1.20& 0.30& 4.297 &1.10E+10& 0.79 &2.05E-02& 6.0& 1.8  & 25.7  & 25.9 &44.89&12:40.1 \nl
 118.000 &4-601.0& 6.56& 0.00& 2.201 &4.85E+09& 0.00 &0.00& 6.0& 1.3  & 27.5  & 27.7 &44.90&11:50.3 \nl
 117.000 &4-455.0& 1.76& 0.10& 1.379 &3.56E+09& 0.49 &3.48E-03& 5.7& 1.6  & 26.5  & 26.8 &44.90&12:19.3 \nl
 119.000 &4-351.0& 2.48& 0.02& 0.626 &2.09E+09& 0.11 &4.26E-04& 5.2& 2.1  & 26.6  & 26.9 &44.93&12:42.7 \nl
 120.000 &4-423.0& 1.76& 0.40& 9.998 &2.11E+10& 0.84 &4.18E-02& 5.8& 2.4  & 25.4  & 25.9 &44.95&12:26.6 \nl
 592.000 &4-458.0& 4.00& 0.10& 3.114 &5.40E+09& 0.49 &4.09E-03& 5.7& 1.4  & 27.3  & 28.0 &44.95&12:17.7 \nl
 123.000 &4-607.0& 1.04& 0.50& 0.618 &1.67E+09& 0.92 &2.47E-03& 6.0& 0.1  & 28.1  & 28.3 &44.96&11:48.8 \nl
 124.000 &\nodata& 1.12& 0.02& 0.499 &2.62E+09& 0.07 &4.55E-04& 3.8& 0.5  & 24.0  & 24.8 &44.99&12:39.7 \nl
 125.000 &4-596.0& 0.48& 0.04& 0.126 &3.41E+08& 0.23 &3.76E-04& 5.6& 2.2  & 26.5  & 26.9 &45.03&11:49.4 \nl
 126.000 &4-515.0& 3.44& 0.10& 0.994 &2.78E+09& 0.50 &1.77E-03& 5.8& 1.1  & 28.1  & 28.2 &45.04&12:3.00 \nl
 127.000 &4-543.0& 1.28& 0.20& 5.356 &1.07E+10& 0.60 &1.69E-02& 6.0& 2.4  & 25.2  & 25.7 &45.05&11:58.1 \nl
 141.120 &4-571.0& 0.72& 0.08& 0.503 &1.59E+09& 0.32 &2.75E-03& 5.0& 3.2  & 24.1  & 24.8 &45.06&11:54.2 \nl
 128.000 &4-431.2& 2.88& 0.00& 0.211 &1.23E+09& 0.00 &0.00& 4.2& 1.6  & 27.3  & 27.7 &45.08&12:22.1 \nl
 1075.00 &4-526.0& 4.72& 0.02& 0.750 &1.82E+09& 0.14 &3.13E-04& 5.9& 1.7  & 28.6  & 28.8 &45.11&12:0.40 \nl
 130.000 &4-573.0& 2.88& 0.00& 0.300 &2.42E+09& 0.00 &0.00& 3.8& 3.3  & 26.7  & 26.9 &45.14&11:50.3 \nl
 1022.00 &\nodata& 1.12& 0.10& 0.603 &1.49E+09& 0.47 &1.23E-03& 5.6& 1.2  & 26.4  & 26.8 &45.14&12:39.4 \nl
 131.000 &4-530.0& 4.56& 0.00& 1.700 &3.66E+09& 0.00 &0.00& 6.0& 3.0  & 27.5  & 27.8 &45.15&11:59.7 \nl
 132.000 &\nodata& 2.00& 0.10& 7.019 &1.23E+10& 0.50 &1.35E-02& 5.8& 3.1  & 24.9  & 25.6 &45.15&12:5.50 \nl
 133.000 &4-329.0& 2.88& 0.00& 1.660 &5.41E+09& 0.00 &0.00& 5.3& 2.1  & 26.1  & 26.3 &45.17&12:44.5 \nl
 1055.00 &\nodata& 2.72& 0.00& 0.373 &9.18E+08& 0.00 &0.00& 6.0& 1.6  & 28.8  & 29.0 &45.22&12:12.5 \nl
 135.000 &4-407.0& 0.56& 0.40& 2.548 &5.26E+09& 0.88 &1.77E-02& 6.0& 1.4  & 25.2  & 25.7 &45.27&12:25.7 \nl
 136.000 &\nodata& 0.40& 0.50& 0.781 &1.93E+09& 0.92 &1.12E-02& 6.0& 2.1  & 26.1  & 26.5 &45.27&11:45.4 \nl
 141.112&4-555.11& 2.96& 0.10& 30.17 &6.34E+10& 0.46 &4.86E-02& 5.5& 3.3  & 23.7  & 24.3 &45.28&11:52.2 \nl
 137.000 &4-342.0& 0.56& 0.08& 0.196 &6.16E+08& 0.45 &7.21E-04& 5.9& 3.2  & 27.3  & 27.4 &45.31&12:38.6 \nl
 138.000 &4-498.0& 1.92& 0.10& 1.694 &4.99E+09& 0.46 &3.66E-03& 5.5& 1.7  & 25.9  & 26.2 &45.32&12:3.50 \nl
 141.111 &4-555.0& 0.80& 0.00& 0.040 &1.06E+10& 0.00 &0.00& 1.2& 0.7  & 21.6  & 22.0 &45.33&11:54.5 \nl
 139.000 &4-395.0& 0.72& 0.20& 1.285 &2.99E+09& 0.60 &7.01E-03& 6.0& 3.8  & 26.3  & 26.7 &45.35&12:26.8 \nl
 140.000 &4-368.0& 1.92& 0.30& 17.11 &3.69E+10& 0.73 &5.40E-02& 5.7& 0.4  & 24.4  & 24.9 &45.36&12:33.7 \nl
 142.000 &4-379.0& 4.32& 0.00& 1.277 &2.18E+09& 0.00 &0.00& 6.0& 1.9  & 28.4  & 29.0 &45.37&12:31.0 \nl
 604.000 &4-559.2& 2.64& 0.00& 0.535 &1.03E+09& 0.00 &0.00& 5.9& 0.0  & 28.0  & 28.5 &45.39&11:47.7 \nl
 141.112 &4-555.1& 3.12& 0.00& 15.28 &5.85E+10& 0.00 &0.00& 4.8& 3.8  & 23.2  & 23.7 &45.41&11:53.2 \nl
 146.000 &4-467.0& 3.20& 0.04& 0.530 &1.56E+09& 0.24 &3.59E-04& 5.7& 2.8  & 28.2  & 28.3 &45.42&12:7.80 \nl
 148.000 &4-557.0& 2.16& 0.00& 1.365 &8.43E+09& 0.00 &0.00& 4.2& 0.4  & 24.6  & 24.9 &45.42&11:49.0 \nl
 150.000 &\nodata& 4.80& 0.20& 10.22 &1.70E+10& 0.60 &1.66E-02& 6.0& 2.1  & 26.9  & 27.5 &45.42&12:2.20 \nl
 147.000 &4-384.2& 1.68& 0.00& 0.158 &4.55E+08& 0.00 &0.00& 5.2& 0.4  & 27.4  & 27.9 &45.43&12:28.3 \nl
 149.000 &4-330.0& 0.96& 0.10& 0.304 &6.29E+08& 0.50 &8.35E-04& 5.8& 2.0  & 27.4  & 28.0 &45.44&12:39.6 \nl
 1077.00 &4-507.0& 3.68& 0.00& 0.000 &6.55E+08& 0.00 &0.00& 6.0& 4.1&\nodata  & 29.9 &45.48&11:57.3 \nl
 607.000 &4-397.0& 3.20& 0.30& 4.452 &8.46E+09& 0.77 &8.76E-03& 5.9& 1.1  & 27.4  & 27.8 &45.49&12:23.6 \nl
 153.000 &4-424.0& 0.56& 0.50& 0.161 &4.57E+08& 0.91 &1.19E-03& 5.9& 0.7  & 27.9  & 28.1 &45.50&12:16.5 \nl
 154.000 &4-487.0& 1.92& 0.02& 0.663 &1.74E+09& 0.15 &3.86E-04& 6.0& 3.9  & 27.5  & 27.6 &45.50&12:2.10 \nl
 155.000 &4-344.0& 1.20& 0.20& 1.659 &3.63E+09& 0.58 &5.99E-03& 5.9& 1.1  & 26.3  & 26.7 &45.51&12:34.6 \nl
 159.000 &4-345.0& 1.36& 0.20& 1.870 &4.84E+09& 0.56 &5.24E-03& 5.8& 2.3  & 26.1  & 26.4 &45.57&12:33.6 \nl
 163.000 &4-389.0& 2.72& 0.08& 4.874 &1.04E+10& 0.42 &7.86E-03& 5.7& 0.4  & 25.8  & 26.3 &45.62&12:25.0 \nl
 165.000 &4-563.0& 4.40& 0.06& 2.847 &2.28E+10& 0.17 &6.89E-03& 3.4& 0.0  & 26.0  & 26.4 &45.64&11:43.1 \nl
 167.000 &4-551.0& 4.00& 0.00& 1.435 &2.68E+09& 0.00 &0.00& 6.0& 2.8  & 27.8  & 28.3 &45.65&11:45.8 \nl
 166.000 &4-307.0& 1.60& 1.00& 534.7 &1.21E+12& 0.97 &1.78E+00& 5.9& 0.6  & 22.5  & 22.9 &45.66&12:42.0 \nl
 141.200 &4-516.0& 1.04& 0.10& 1.329 &8.95E+09& 0.31 &5.10E-03& 4.1& 0.8  & 23.2  & 23.6 &45.66&11:54.0 \nl
 170.000 &\nodata& 0.32& 0.50& 0.028 &6.78E+07& 0.81 &3.84E-04& 5.3& 0.1  & 27.1  & 27.8 &45.73&12:27.1 \nl
 171.000 &4-323.0& 2.80& 0.00& 0.295 &1.04E+09& 0.00 &0.00& 5.6& 3.1  & 28.3  & 28.2 &45.73&12:35.9 \nl
 172.000 &4-497.0& 2.88& 0.00& 4.303 &1.16E+10& 0.00 &0.00& 5.4& 2.3  & 25.2  & 25.6 &45.74&11:57.3 \nl
 174.000 &4-478.0& 2.64& 0.10& 0.487 &1.74E+09& 0.47 &9.80E-04& 5.6& 0.1  & 28.2  & 28.1 &45.77&11:59.1 \nl
 177.000 &4-520.0& 0.00& 0.00& 0.000 &2.15E+03& 0.00 &0.00& 4.4& 1.3  & 25.1  & 25.4 &45.79&11:50.6 \nl
 175.000 &4-419.0& 1.92& 0.20& 1.719 &3.65E+09& 0.60 &4.02E-03& 6.0& 2.8  & 27.0  & 27.4 &45.79&12:13.7 \nl
 176.000 &4-387.0& 0.56& 0.30& 0.374 &8.69E+08& 0.77 &2.36E-03& 5.9& 2.0  & 26.7  & 27.2 &45.80&12:22.5 \nl
 1066.00 &4-464.0& 0.56& 0.02& 0.040 &1.24E+08& 0.14 &4.74E-05& 5.8& 1.9  & 28.4  & 28.5 &45.84&12:1.10 \nl
 184.000 &4-473.0& 5.52& 0.00& 4.507 &9.08E+09& 0.00 &0.00& 6.0& 0.4  & 26.6  & 26.9 &45.88&11:58.2 \nl
 625.000 &4-441.0& 3.20& 0.20& 1.227 &3.14E+09& 0.58 &1.82E-03& 5.9& 1.1  & 28.3  & 28.4 &45.91&12:5.60 \nl
 1049.00 &4-414.0& 0.16& 0.10& 0.009 &1.73E+07& 0.54 &2.47E-04& 6.0& 1.4  & 29.4  & 30.1 &45.92&12:12.7 \nl
 1029.00 &4-343.0& 0.48& 0.50& 0.382 &4.15E+08& 0.92 &3.81E-03& 6.0& 0.8  & 27.4  & 28.6 &45.92&12:27.9 \nl
 188.000 &4-514.0& 1.20& 0.08& 1.490 &3.71E+09& 0.43 &4.16E-03& 5.8& 1.2  & 25.8  & 26.2 &45.94&11:49.5 \nl
 187.000 &4-326.0& 1.92& 0.10& 1.005 &2.93E+09& 0.50 &2.11E-03& 5.8& 1.3  & 26.9  & 27.0 &45.95&12:31.5 \nl
 186.000 &4-305.0& 0.96& 0.40& 4.937 &1.03E+10& 0.88 &2.21E-02& 6.0& 5.4  & 25.4  & 25.9 &45.95&12:38.3 \nl
 189.000 &4-460.0& 0.56& 0.40& 7.682 &1.54E+10& 0.86 &5.39E-02& 5.9& 1.5  & 23.6  & 24.2 &45.96&12:1.40 \nl
 190.000 &4-391.0& 0.08& 0.08& 0.000 &5.79E+06& 0.45 &0.00& 5.9& 2.8&\nodata  & 28.7 &46.01&12:17.7 \nl
 191.000 &4-353.0& 2.64& 0.06& 1.700 &3.58E+09& 0.35 &2.36E-03& 5.8& 0.2  & 26.9  & 27.4 &46.04&12:23.8 \nl
 193.000 &\nodata& 1.68& 0.02& 0.077 &4.04E+08& 0.07 &5.42E-05& 4.0& 5.8  & 27.3  & 27.9 &46.09&12:26.7 \nl
 628.000 &4-293.0& 0.96& 0.40& 1.173 &1.76E+09& 0.88 &5.25E-03& 6.0& 0.1  & 27.1  & 28.0 &46.10&12:36.6 \nl
 195.000 &4-331.0& 3.36& 0.02& 1.433 &2.71E+09& 0.12 &7.60E-04& 5.4& 0.6  & 26.9  & 27.6 &46.14&12:27.4 \nl
 198.200 &4-522.0& 1.12& 0.30& 7.191 &1.54E+10& 0.79 &2.12E-02& 6.0& 2.6  & 25.0  & 25.5 &46.15&11:45.0 \nl
 197.000 &4-299.0& 4.24& 0.02& 1.400 &3.67E+09& 0.11 &4.72E-04& 5.3& 1.9  & 27.3  & 27.7 &46.16&12:34.6 \nl
 196.000 &4-285.0& 2.16& 0.20& 0.877 &2.52E+09& 0.48 &1.76E-03& 5.3& 1.1  & 26.8  & 27.2 &46.17&12:36.8 \nl
 632.000 &4-396.2& 3.20& 0.20& 1.538 &3.02E+09& 0.58 &2.28E-03& 5.9& 1.9  & 28.0  & 28.4 &46.17&12:12.4 \nl
 198.100 &\nodata& 1.04& 0.08& 1.250 &2.40E+10& 0.17 &7.67E-03& 2.7& 1.4  & 21.2  & 21.9 &46.18&11:42.1 \nl
 633.000 &4-362.0& 0.48& 0.30& 0.199 &4.34E+08& 0.79 &1.71E-03& 6.0& 1.9  & 27.9  & 28.3 &46.19&12:20.3 \nl
 200.000 &4-465.0& 0.48& 0.50& 0.158 &3.94E+08& 0.92 &1.57E-03& 6.0& 1.1  & 28.3  & 28.6 &46.20&11:55.2 \nl
 199.000 &4-294.0& 3.76& 0.08& 1.979 &4.31E+09& 0.38 &2.65E-03& 5.4& 1.9  & 27.1  & 27.6 &46.21&12:35.2 \nl
 201.100 &4-322.0& 1.68& 0.30& 23.32 &5.57E+10& 0.75 &5.64E-02& 5.8& 3.1  & 24.0  & 24.3 &46.21&12:28.4 \nl
 201.000 &\nodata& 0.00& 0.00& 0.000 &2.69E+03& 0.00 &0.00& 4.1& 2.9  & 24.3  & 24.9 &46.23&12:29.1 \nl
 204.000 &4-448.0& 0.48& 0.50& 3.561 &6.00E+09& 0.92 &3.55E-02& 6.0& 4.4  & 24.7  & 25.4 &46.27&11:59.8 \nl
 203.000 &4-341.0& 3.36& 0.08& 1.558 &4.24E+09& 0.42 &2.52E-03& 5.7& 0.3  & 27.4  & 27.6 &46.27&12:22.8 \nl
 202.000 &4-300.0& 0.56& 0.30& 1.424 &3.60E+09& 0.73 &9.14E-03& 5.7& 3.3  & 24.7  & 25.1 &46.27&12:33.5 \nl
 207.000 &4-327.0& 1.84& 0.30& 9.414 &2.03E+10& 0.79 &3.17E-02& 6.0& 0.5  & 25.5  & 25.9 &46.36&12:24.8 \nl
 640.000 &4-390.0& 1.76& 0.20& 0.907 &1.55E+09& 0.60 &2.54E-03& 6.0& 0.7  & 27.9  & 28.5 &46.37&12:11.6 \nl
 208.000 &4-372.0& 0.48& 0.50& 0.987 &2.20E+09& 0.92 &9.83E-03& 6.0& 2.1  & 26.2  & 26.6 &46.39&12:15.8 \nl
 209.000 &\nodata& 0.88& 0.02& 0.548 &1.17E+09& 0.11 &5.41E-04& 5.2& 0.2  & 24.9  & 25.8 &46.41&12:4.60 \nl
 210.000 &4-489.0& 1.84& 0.08& 0.564 &1.53E+09& 0.41 &1.13E-03& 5.6& 1.6  & 27.1  & 27.4 &46.42&11:46.5 \nl
 647.000 &4-443.0& 1.76& 0.10& 0.329 &7.42E+08& 0.46 &8.52E-04& 5.5& 1.4  & 27.6  & 28.2 &46.45&11:56.3 \nl
 645.000 &4-262.2& 5.52& 0.00& 1.483 &3.61E+09& 0.00 &0.00& 6.0& 2.7  & 28.1  & 28.2 &46.45&12:37.5 \nl
 211.000 &4-303.0& 1.92& 0.30& 1.606 &4.02E+09& 0.69 &5.26E-03& 5.5& 2.4  & 26.6  & 27.0 &46.49&12:27.9 \nl
 212.000 &4-471.0& 0.32& 0.00& 0.009 &2.74E+09& 0.00 &0.00& 1.2& 6.5  & 21.2  & 21.5 &46.51&11:51.4 \nl
 213.000 &4-488.0& 2.48& 0.02& 3.169 &7.48E+09& 0.14 &1.97E-03& 5.9& 5.1  & 25.8  & 26.1 &46.53&11:45.6 \nl
 1054.00 &\nodata& 0.48& 0.40& 0.065 &1.88E+08& 0.86 &6.29E-04& 5.9& 0.7  & 28.5  & 28.6 &46.54&12:6.10 \nl
 214.000 &4-416.0& 0.48& 0.30& 3.081 &7.30E+09& 0.77 &2.64E-02& 5.9& 1.7  & 24.2  & 24.6 &46.55&12:3.10 \nl
 216.000 &4-434.0& 0.16& 0.40& 0.227 &5.36E+08& 0.88 &9.67E-03& 6.0& 3.1  & 25.5  & 25.9 &46.58&11:57.2 \nl
 215.000 &4-350.0& 1.20& 0.00& 0.079 &6.38E+08& 0.00 &0.00& 3.8& 1.2  & 26.2  & 26.5 &46.59&12:15.5 \nl
 218.000 &4-412.0& 3.20& 0.06& 0.946 &2.18E+09& 0.32 &9.02E-04& 5.5& 2.6  & 27.5  & 27.9 &46.60&12:1.90 \nl
 220.000 &4-475.0& 3.68& 0.02& 3.183 &8.01E+09& 0.12 &1.40E-03& 5.4& 4.4  & 26.2  & 26.6 &46.65&11:46.3 \nl
 219.000 &4-280.0& 3.84& 0.02& 2.027 &5.61E+09& 0.15 &7.98E-04& 6.0& 8.3  & 27.2  & 27.2 &46.67&12:30.1 \nl
 221.000 &4-411.0& 0.16& 0.02& 0.004 &2.99E+07& 0.14 &3.35E-05& 5.8& 1.4  & 28.8  & 28.2 &46.67&12:0.80 \nl
 222.000 &4-442.0& 0.64& 0.10& 0.232 &9.41E+08& 0.38 &9.93E-04& 5.0& 0.6  & 24.8  & 25.2 &46.67&11:54.0 \nl
 652.000 &4-388.0& 2.88& 0.02& 0.760 &1.61E+09& 0.13 &3.47E-04& 5.7& 1.3  & 27.5  & 28.0 &46.69&12:7.10 \nl
 225.000 &4-336.0& 2.32& 0.02& 0.786 &2.40E+09& 0.11 &3.37E-04& 5.2& 3.1  & 26.2  & 26.6 &46.73&12:16.1 \nl
 226.000 &4-334.0& 0.48& 0.50& 0.709 &1.82E+09& 0.92 &7.07E-03& 6.0& 5.2  & 26.6  & 26.9 &46.73&12:17.6 \nl
 228.000 &4-388.2& 2.96& 0.04& 0.472 &1.48E+09& 0.19 &4.36E-04& 5.1& 0.0  & 27.4  & 27.8 &46.77&12:6.80 \nl
 230.000 &4-432.0& 0.48& 0.30& 0.183 &4.27E+08& 0.75 &1.58E-03& 5.8& 1.3  & 26.9  & 27.3 &46.80&11:53.7 \nl
 1018.00 &\nodata& 0.32& 0.40& 0.015 &3.83E+07& 0.71 &2.13E-04& 5.1& 3.3  & 27.2  & 28.1 &46.80&12:34.8 \nl
 231.000 &4-382.2& 2.88& 0.04& 0.924 &2.42E+09& 0.22 &8.02E-04& 5.4& 1.3  & 27.1  & 27.4 &46.81&12:5.80 \nl
 232.000 &4-360.0& 0.00& 0.00& 0.000 &2.50E+02& 0.00 &0.00& 4.7& 0.8  & 28.0  & 28.1 &46.82&12:9.40 \nl
 235.000 &4-289.0& 2.88& 0.00& 6.719 &1.53E+10& 0.00 &0.00& 5.7& 1.6  & 25.1  & 25.5 &46.93&12:26.1 \nl
 236.000 &4-295.0& 1.12& 0.30& 1.636 &3.94E+09& 0.79 &4.82E-03& 6.0& 6.6  & 26.7  & 27.0 &46.94&12:23.1 \nl
 237.000 &4-382.0& 0.08& 0.60& 0.269 &5.45E+08& 0.95 &4.91E-02& 6.0& 5.5  & 24.2  & 24.8 &46.95&12:5.40 \nl
 1027.00 &4-295.2& 0.96& 0.30& 0.778 &1.62E+09& 0.79 &3.14E-03& 6.0& 1.1  & 27.3  & 27.8 &46.96&12:23.4 \nl
 238.000 &4-415.0& 0.48& 0.20& 0.460 &1.08E+09& 0.56 &2.99E-03& 5.8& 2.9  & 25.8  & 26.3 &47.00&11:56.3 \nl
 1026.00 &4-266.0& 0.48& 0.20& 0.130 &2.62E+08& 0.60 &8.44E-04& 6.0& 2.3  & 28.1  & 28.6 &47.02&12:26.8 \nl
 240.000 &4-332.0& 0.48& 0.20& 1.440 &3.34E+09& 0.58 &9.31E-03& 5.9& 0.9  & 25.0  & 25.4 &47.08&12:12.5 \nl
 662.000 &\nodata& 0.24& 0.70& 0.005 &1.96E+07& 0.80 &1.68E-04& 5.0& 0.1  & 27.7  & 28.2 &47.14&12:0.60 \nl
 663.000 &4-216.2& 3.92& 0.00& 1.496 &2.38E+09& 0.00 &0.00& 6.0& 3.2  & 27.8  & 28.4 &47.18&12:33.1 \nl
 241.000 &4-335.0& 0.56& 0.20& 0.371 &7.86E+08& 0.60 &1.76E-03& 6.0& 1.4  & 27.1  & 27.6 &47.22&12:8.60 \nl
 242.000 &4-332.2& 0.56& 0.10& 0.546 &1.52E+09& 0.52 &2.32E-03& 5.9& 1.5  & 26.2  & 26.4 &47.23&12:12.6 \nl
 243.000 &4-385.0& 0.16& 0.40& 0.167 &3.86E+08& 0.88 &7.12E-03& 6.0& 2.7  & 25.9  & 26.3 &47.23&11:59.0 \nl
 1050.00 &\nodata& 1.84& 0.02& 0.106 &2.99E+08& 0.11 &7.36E-05& 5.2& 4.1  & 28.0  & 28.5 &47.23&12:5.10 \nl
 244.000 &4-232.0& 0.40& 0.40& 7.773 &1.54E+10& 0.86 &1.07E-01& 5.9& 3.5  & 23.0  & 23.6 &47.29&12:30.7 \nl
 244.200 &4-232.2& 0.48& 0.30& 0.395 &9.33E+08& 0.75 &3.41E-03& 5.8& 2.9  & 26.1  & 26.6 &47.33&12:29.0 \nl
 245.000 &4-298.0& 2.64& 0.04& 1.804 &4.24E+09& 0.27 &1.82E-03& 6.0& 6.9  & 27.1  & 27.3 &47.34&12:15.8 \nl
 246.000 &4-356.0& 1.12& 0.06& 0.274 &8.01E+08& 0.32 &3.93E-04& 5.5& 0.3  & 26.9  & 27.2 &47.41&12:0.60 \nl
 247.000 &4-319.0& 0.72& 0.40& 1.988 &4.32E+09& 0.88 &1.59E-02& 6.0& 3.2  & 26.0  & 26.4 &47.46&12:8.40 \nl
 248.000 &\nodata& 5.76& 0.00& 2.331 &5.37E+09& 0.00 &0.00& 5.3& 0.3  & 27.2  & 27.7 &47.46&11:59.9 \nl
 249.000 &4-315.0& 1.76& 0.00& 1.824 &5.11E+09& 0.00 &0.00& 5.2& 1.9  & 24.9  & 25.4 &47.48&12:11.2 \nl
 250.000 &4-291.0& 0.96& 0.60& 1.686 &4.51E+09& 0.95 &8.12E-03& 6.0& 1.9  & 27.0  & 27.2 &47.51&12:14.1 \nl
 251.000 &4-346.0& 0.72& 0.04& 0.396 &1.39E+09& 0.18 &1.24E-03& 5.0& 8.3  & 24.3  & 24.9 &47.51&12:2.60 \nl
 1092.00 &\nodata& 0.48& 0.50& 0.132 &4.13E+08& 0.92 &1.32E-03& 6.0& 1.4  & 28.4  & 28.4 &47.64&12:31.6 \nl
 255.000 &4-301.2& 1.36& 0.00& 0.255 &4.67E+08& 0.00 &0.00& 5.6& 1.2  & 27.4  & 28.1 &47.66&12:8.90 \nl
 256.000 &4-283.0& 2.88& 0.00& 0.717 &2.38E+09& 0.00 &0.00& 5.6& 6.1  & 27.3  & 27.3 &47.67&12:12.5 \nl
 257.000 &4-231.0& 0.64& 0.10& 0.076 &2.19E+08& 0.44 &2.62E-04& 5.4& 0.6  & 27.0  & 27.4 &47.67&12:22.1 \nl
 258.000 &4-308.0& 2.16& 0.00& 0.642 &1.69E+09& 0.00 &0.00& 5.5& 3.4  & 26.7  & 27.1 &47.67&12:8.20 \nl
 259.000 &4-247.0& 0.40& 0.00& 0.019 &6.83E+07& 0.00 &0.00& 5.5& 2.5  & 28.1  & 28.2 &47.69&12:19.3 \nl
 260.000 &\nodata& 0.08& 1.00& 0.000 &3.51E+06& 0.88 &1.74E-04& 5.0& 0.6  & 28.1  & 28.1 &47.81&12:11.0 \nl
 261.000 &4-212.0& 1.12& 0.06& 0.963 &2.38E+09& 0.33 &1.36E-03& 5.6& 0.2  & 25.7  & 26.2 &47.82&12:24.2 \nl
 1040.00 &\nodata& 5.44& 0.00& 1.374 &2.16E+09& 0.00 &0.00& 6.0& 0.7  & 28.1  & 28.7 &47.83&12:4.50 \nl
 264.000 &4-304.0& 0.72& 0.50& 1.385 &3.68E+09& 0.91 &1.18E-02& 5.9& 1.1  & 26.1  & 26.3 &47.86&12:5.40 \nl
 265.000 &4-174.0& 0.00& 0.20& 0.000 &7.24E+02& 0.43 &0.00& 5.0& 0.3  & 27.4  & 27.5 &47.89&12:29.5 \nl
 680.000 &\nodata& 0.00& 0.00& 0.000 &3.18E+02& 0.00 &0.00& 5.4& 0.9  & 29.2  & 29.0 &47.91&12:18.5 \nl
 267.000 &4-314.0& 4.80& 0.00& 2.280 &4.97E+09& 0.00 &0.00& 6.0& 0.7  & 27.3  & 27.5 &48.00&12:0.80 \nl
 1025.00 &4-198.0& 0.96& 0.04& 0.000 &2.53E+08& 0.27 &0.00& 6.0& 2.7&\nodata  & 30.8 &48.06&12:22.5 \nl
 277.120 &4-260.0& 0.56& 0.30& 3.895 &8.90E+09& 0.77 &2.45E-02& 5.9& 2.7  & 24.2  & 24.6 &48.12&12:14.9 \nl
 269.000 &4-265.0& 0.00& 0.40& 0.000 &9.87E+02& 0.82 &0.00& 5.7& 1.0  & 28.3  & 28.6 &48.15&12:8.40 \nl
 277.100&4-260.11& 0.88& 0.00& 0.392 &2.90E+09& 0.00 &0.00& 3.2& 0.4  & 23.1  & 23.9 &48.25&12:13.9 \nl
 273.000 &4-170.0& 0.48& 0.30& 0.114 &3.17E+08& 0.77 &9.85E-04& 5.9& 2.1  & 27.7  & 27.9 &48.31&12:22.8 \nl
 685.000 &4-158.0& 3.52& 0.06& 0.798 &1.86E+09& 0.35 &9.48E-04& 5.8& 4.0  & 28.1  & 28.3 &48.35&12:23.9 \nl
 274.000 &4-200.0& 4.72& 0.04& 5.725 &1.50E+10& 0.20 &4.83E-03& 5.2& 9.7  & 26.0  & 26.5 &48.37&12:17.3 \nl
 275.000 &4-237.0& 0.96& 0.20& 0.401 &1.02E+09& 0.58 &1.22E-03& 5.9& 1.9  & 27.5  & 27.8 &48.41&12:9.00 \nl
 276.000 &4-135.0& 0.08& 0.70& 0.016 &4.68E+07& 0.95 &2.93E-03& 6.0& 0.6  & 27.4  & 27.5 &48.43&12:27.0 \nl
 278.000 &4-228.0& 1.68& 0.10& 0.253 &7.02E+08& 0.49 &4.14E-04& 5.7& 1.3  & 28.2  & 28.4 &48.50&12:8.40 \nl
 279.000 &\nodata& 1.36& 0.20& 1.910 &3.98E+09& 0.58 &5.33E-03& 5.9& 3.3  & 26.3  & 26.8 &48.57&12:27.0 \nl
 280.000 &4-257.0& 0.56& 0.20& 0.701 &1.98E+09& 0.53 &3.40E-03& 5.6& 3.8  & 25.0  & 25.4 &48.58&12:3.90 \nl
 281.000 &4-229.0& 0.80& 0.20& 1.190 &3.29E+09& 0.55 &5.30E-03& 5.7& 2.0  & 25.4  & 25.7 &48.62&12:7.80 \nl
 277.211 &4-186.0& 1.84& 0.70& 375.2 &1.07E+12& 0.95 &1.51E+00& 6.0& 2.8  & 22.4  & 22.9 &48.62&12:15.8 \nl
 277.220 &4-260.2& 0.48& 0.20& 0.746 &1.62E+09& 0.51 &5.00E-03& 5.5& 1.7  & 24.5  & 25.1 &48.65&12:14.2 \nl
 282.000 &4-111.0& 0.00& 0.20& 0.000 &7.02E+02& 0.49 &0.00& 5.4& 10.  & 28.8  & 28.7 &48.67&12:26.3 \nl
 277.212 &4-169.0& 5.04& 0.08& 16.84 &1.19E+11& 0.22 &3.74E-02& 3.5& 0.0  & 24.6  & 25.1 &48.71&12:16.7 \nl
 1023.00 &4-143.0& 2.64& 0.00& -0.00 &4.74E+08& 0.00 &\nodata& 6.0& 3.8&\nodata&\nodata &48.74&12:20.5 \nl
 283.000 &4-154.0& 0.48& 0.40& 0.815 &1.91E+09& 0.84 &7.88E-03& 5.8& 6.8  & 25.4  & 25.8 &48.75&12:19.1 \nl
 690.000 &4-203.2& 1.52& 0.10& 0.590 &1.08E+09& 0.52 &1.18E-03& 5.9& 1.6  & 27.6  & 28.2 &48.82&12:8.40 \nl
 284.000 &4-183.0& 0.32& 0.30& 0.003 &2.18E+07& 0.55 &5.49E-05& 4.6& 0.2  & 27.8  & 28.0 &48.84&12:12.0 \nl
 693.000 &\nodata& 4.88& 0.00& 1.045 &2.04E+09& 0.00 &0.00& 6.0& 0.2  & 28.2  & 28.6 &48.86&12:16.8 \nl
 285.000 &4-128.0& 1.92& 0.10& 1.019 &1.91E+09& 0.54 &2.14E-03& 6.0& 3.5  & 27.2  & 27.7 &48.88&12:20.2 \nl
 286.000 &4-199.0& 0.00& 0.08& 0.000 &3.08E+02& 0.25 &0.00& 4.0& 0.7  & 26.9  & 27.4 &48.89&12:8.70 \nl
 288.100 &4-203.0& 0.48& 0.50& 2.006 &4.64E+09& 0.92 &2.00E-02& 6.0& 1.5  & 25.4  & 25.7 &48.91&12:8.00 \nl
 287.000 &4-148.0& 4.64& 0.00& 1.558 &2.31E+09& 0.00 &0.00& 6.0& 1.5  & 27.9  & 28.6 &48.92&12:16.7 \nl
 1019.00 &4-105.0& 2.16& 0.00& 0.279 &5.38E+08& 0.00 &0.00& 5.9& 0.2  & 28.3  & 28.8 &48.97&12:23.2 \nl
 288.200&4-203.12& 2.80& 0.02& 0.790 &2.88E+09& 0.10 &4.41E-04& 5.1& 3.7  & 26.5  & 26.8 &49.01&12:8.20 \nl
 289.000 &\nodata& 0.72& 0.08& 1.699 &7.07E+09& 0.32 &9.29E-03& 5.0& 2.0  & 22.4  & 23.2 &49.06&12:21.3 \nl
 696.000 &4-124.0& 0.32& 0.00& 0.003 &1.47E+07& 0.00 &0.00& 5.0& 2.9  & 28.5  & 28.9 &49.18&12:16.1 \nl
 290.000  &4-99.0& 0.56& 0.08& 0.111 &2.76E+08& 0.43 &4.11E-04& 5.8& 4.4  & 27.5  & 27.8 &49.28&12:20.0 \nl
 292.000 &4-131.0& 0.72& 0.30& 3.471 &7.08E+09& 0.77 &2.51E-02& 5.9& 2.1  & 24.8  & 25.3 &49.33&12:14.5 \nl
 293.000 &\nodata& 2.64& 0.10& 2.307 &4.71E+09& 0.47 &4.64E-03& 5.6& 3.0  & 26.5  & 27.1 &49.47&12:22.5 \nl
 294.000 &4-119.0& 0.80& 0.00& 0.129 &3.46E+08& 0.00 &0.00& 5.6& 3.4  & 27.0  & 27.4 &49.51&12:12.1 \nl
 295.100  &4-85.0& 0.96& 0.30& 10.28 &2.19E+10& 0.77 &4.16E-02& 5.9& 3.1  & 24.1  & 24.6 &49.52&12:20.1 \nl
 296.000 &4-109.0& 1.92& 0.20& 18.18 &4.11E+10& 0.60 &4.25E-02& 6.0& 4.1  & 24.3  & 24.7 &49.59&12:12.7 \nl
 298.000  &4-95.0& 2.08& 0.00& 0.162 &9.74E+08& 0.00 &0.00& 4.3& 2.1  & 26.9  & 27.2 &49.60&12:15.3 \nl
 295.200  &4-85.2& 2.72& 0.00& 5.961 &1.65E+10& 0.00 &0.00& 5.3& 1.3  & 24.6  & 25.0 &49.68&12:19.7 \nl
 301.000 &\nodata& 0.80& 0.00& 0.081 &6.13E+09& 0.00 &0.00& 1.7& 2.4  & 22.0  & 22.6 &50.14&12:17.4 
\enddata
\tablenotetext{a}{This is the fraction of the luminosity removed by extinction and re-emitted in the mid and far infrared}
\tablenotetext{b}{The selected template number between 1.0 (early-cold) and 6.0 (late-hot).}
\tablenotetext{c}{The total AB magnitude in the F160W filter.}
\tablenotetext{d}{The F160W magnitude in an $0.6\arcsec$ diameter aperture.}
\end{deluxetable}
\clearpage

\begin{deluxetable}{cccccccccccc}
\scriptsize
\tablecaption{The ten largest contributors to the star formation rate in each
redshift bin. The percentage of the total star formation rate for the bin is
listed.
\label{tab2}}
\tablewidth{0pt}
\tablehead{
\colhead{Redshift Bin} & \colhead{} & \multicolumn{10}{c} {Object Rank}}
\startdata
 & &1&2&3&4&5&6&7&8&9&10 \\ \hline \nl
0.5 - 1.5 & & 32.4& 10.3& 9.04& 4.93& 3.68& 3.45& 2.57& 2.36& 2.06& 1.86\nl  & ID & 49.0000&68.0000&97.0000&295.100&189.000&198.200&127.000&186.000&901.000&277.120 \\ \hline \nl
1.5 - 2.5& & 46.8& 32.8& 7.04& 2.04& 1.86& 1.59& 1.49& 0.87& 0.82& 0.61\nl
 & ID &166.0&277.211&102.000&201.10&61.0&296.0&140.0&120.0&207.0&132.0 \\ \hline \nl
2.5 - 3.5& & 25.8& 13.1& 5.76& 5.11& 5.04& 4.18& 3.81& 3.69& 3.08& 2.57\nl
 & ID &141.112&141.112&235.0&295.20&81.10&163.0&607.0&172.0&1084.0&48.0 \\ \hline \nl
3.5 - 4.5& & 38.7& 5.81& 5.68& 5.19& 4.94& 3.70& 3.61& 3.33& 3.14& 2.96\nl
 & ID &103.0&220.0&592.0&165.0&2.0&219.0&199.0&528.0&1062.0&34.0 \\ \hline \nl
4.5 - 5.5& & 30.5& 18.5& 11.1& 10.4& 10.3& 4.13& 3.08& 2.82& 2.49& 1.89\nl
 & ID &277.212&150.0&96.0&92.0&274.0&267.0&131.0&287.0&1040.0&693.0 \\ \hline \nl
5.5 - 6.5& & 32.8& 21.8& 17.4& 17.0& 10.8&\nodata&\nodata&\nodata&\nodata&\nodata\nl
 & ID &184.0&586.0&107.0&248.0&645.0&\nodata&\nodata&\nodata&\nodata&\nodata
\enddata 
\end{deluxetable}
\clearpage

\begin{deluxetable}{cccc}
\footnotesize
\tablecaption{Fraction of Missing Luminosity as a Function of Redshift \label{tab3}}
\tablewidth{0pt}
\tablehead{
\colhead{Redshift} & \colhead{Const. L$^*$ Fraction} & \colhead{Evolv. L$^*$ Fraction} & \colhead{Ext. Cor. SFR Inten. Dist}}
\startdata
1.0 &0.12 &0.15 &0.00 \nl
2.0 &0.17 &0.29 &0.31 \nl
3.0 &0.24 &0.46 &0.20 \nl
4.0 &0.33 &0.67 &0.57 \nl
5.0 &0.38 &0.79 &0.46 \nl
6.0 &0.42 &0.87 &0.31 
\enddata 
\end{deluxetable}
\clearpage

\begin{deluxetable}{cccc}
\footnotesize
\tablecaption{The log of the star formation rates at different correction levels
in solar masses per year. 
\label{tab4}}
\tablewidth{0pt}
\tablehead{
\colhead{Redshift} & \colhead{Uncorrected} & \colhead{Extinction Corrected} & \colhead{Ext. and Incompl. Corrected}}
\startdata
1.0 &-1.9 &-0.72 &-0.72 \nl
2.0 &-1.5 &-0.19 &-0.03 \nl
3.0 &-1.4 &-1.2 &-1.14 \nl
4.0 &-1.8 &-1.6 &-1.22 \nl
5.0 &-1.8 &-1.6 &-1.31 \nl
6.0 &-2.2 &-2.2 &-2.01  
\enddata
\end{deluxetable}

\clearpage

\begin{deluxetable}{ccccccc}
\footnotesize
\tablecaption{Sources of Numerical Variances 
\label{tab5}}
\tablewidth{0pt}
\tablehead{
\colhead{Statistic} & \colhead{z = 1} & \colhead{z = 2} & \colhead{z = 3} & \colhead{z = 4} & \colhead{z = 5} & \colhead{z = 6}}
\startdata
Number of Sources &81 &52 &40 &22 &13 &5 \nl
$1/\sqrt{N}$ &0.11 &0.14 &0.16 &0.21 &0.28 &0.45 \nl
I$_2$ &0.19 &0.22 &0.26 &0.31 &0.36 &0.42 \nl
$\sigma_N/N$ &0.45 &0.49 &0.54 &0.60 &0.66 &0.79 \nl
Bootstrap fractional error &0.42 &0.54 &0.28 &0.38 &0.29 &0.16 \nl
Quadrature sum of rows 4 and 5 &0.62 &0.73 &0.61 &0.71 &0.72 &0.81 \nl
$16 \%$ Confidence fraction &0.35 &0.51 &0.2 &0.08 &0.4 &0.8 \nl
$84 \%$ Confidence fraction &0.38 &0.62 &0.50 &0.70 &2.70 &1.90 \nl 
Lower Template Degen. error &0.5 &0.5 &0.5 &0.5 &0.5 &0.5 \nl
Upper Template Degen. error &1.0 &1.0 &1.0 &1.0 &1.0 &1.0 \nl
Lower Total error &0.88 &0.93 &0.84 &0.87 &0.92 &0.98 \nl
Upper Total error &1.24 &1.38 &1.27 &1.41 &2.97 &2.29  
\enddata 
\end{deluxetable}

\clearpage

\begin{deluxetable}{ccc}
\tablecaption{Comparison of Observed and Model Star Formation Rates
\label{tab6}}
\tablewidth{0pt}
\tablehead{
\colhead{Values\tablenotemark{a}} & \colhead{NICMOS} & \colhead{Weinberg\tablenotemark{b}}}
\startdata
Log($SFR_{25}$) &0.5 &0.6 \nl
Log($SFR_{50}/SFR_{25}$) &-0.2 &-0.3 \nl
Log($SFR_{75}/SFR_{25}$) &-0.4 &-1.0 \nl
\enddata 
\tablenotetext{a}{See the text for an explanation of the values}
\tablenotetext{b}{These values taken from Table 2 in \cite{wein99}}
\end{deluxetable}
\clearpage

\end{document}